\documentstyle[aps,eqsecnum,multicol,psfig]{revtex}
\begin{document}
\newcommand{\beq}{\begin{equation}}
\newcommand{\eeq}{\end{equation}}
\newcommand{\bea}{\begin{eqnarray}}
\newcommand{\eea}{\end{eqnarray}}
\newcommand{\bml}{\begin{mathletters}}
\newcommand{\eml}{\end{mathletters}}
\newcommand{\ie}{{i.e.}}
\newcommand{\eg}{{e.g.}}
\newcommand{\rhs}{{r.h.s.}}
\newcommand{\nn}{\nonumber \\}
\def\real{{\rm Re\,}}           
\def\imag{{\rm Im\,}}           
\newcommand{\eqbreak}{
\end{multicols}
\widetext
\noindent
\rule{.48\linewidth}{.1mm}\rule{.1mm}{.1cm}
}
\newcommand{\eqresume}{
\noindent
\rule{.52\linewidth}{.0mm}\rule[-.1cm]{.1mm}{.1cm}\rule{.48\linewidth}{.1mm}
\begin{multicols}{2}
\narrowtext
}

\title{
The amorphous solid state: a locally stable thermodynamic phase of
randomly constrained systems
} 

\author{Horacio E.~Castillo\rlap,$^{(a)}$ 
Paul M.~Goldbart$^{(b)}$ 
and Annette Zippelius$^{(c)}$}

\address{
$^{(a)}$ CNRS-Laboratoire de Physique Th{\'e}orique de l'Ecole
Normale Sup{\'e}rieure,
24 rue Lhomond, 
75231 Paris CEDEX 05, France\\
$^{(b)}$ Department of Physics, University of Illinois at
Urbana-Champaign, 1110 West Green Street, 
Urbana, IL 61801-3080, USA\\
$^{(c)}$ Institut f{\"u}r Theoretische Physik,
Georg August Universit{\"a}t, 
D 37073 G{\"o}ttingen, Germany
}
\date{May 21, 1999}
\maketitle

\begin{abstract}
The question of the local stability of the (replica-symmetric)
amorphous solid state is addressed for a class of systems undergoing a
continuous liquid to amorphous-solid phase transition driven by the
effect of random constraints. The Hessian matrix, associated with
infinitesimal fluctuations around the stationary point corresponding
to the amorphous solid state, is obtained. The eigenvalues of this
Hessian matrix are all shown to be strictly positive near the transition,
except for one---the zero mode associated with the spontaneously broken
continuous translational symmetry of the system. Thus the local
stability of the amorphous solid state is established.
\end{abstract}

\pacs{64.60.Ak, 82.70.Gg, 61.43.-j}
%
%

\begin{multicols}{2}

\section{Introduction}
\label{SEC:introduction}

In recent years, a theoretical approach has been developed for the
problem of the liquid-amorphous solid phase transition in systems of
randomly crosslinked flexible linear
macromolecules~\cite{REF:prl_1987,REF:PMGandAZprl,REF:epl,REF:cross}. This
approach starts from a semi-microscopic model for the macromolecules,
and takes into account explicitly both the thermal fluctuations at
nonzero temperature and the quenched disorder due to the random nature
of the crosslinking. It is based on the Deam-Edwards
formulation of the statistical mechanics of polymer
networks~\cite{REF:DeamEd}, and borrows some concepts and techniques that
have in the past been used to study the problem of spin
glasses~\cite{REF:MPVbook}. 
In the framework of this approach, it is possible to define an order
parameter that probes random static particle-density fluctuations,
and is thus able to detect the transition between the liquid and the
amorphous solid state~\cite{REF:prl_1987,REF:cross}.

Along the way, it has been recognized that there exists a class of
systems (including, e.g., end-linked flexible and stiff
polymers~\cite{REF:end}, and crosslinked higher-dimensional
manifolds~\cite{REF:manifolds}) that display identical transitions,
and a general Landau theory was formulated to describe this
transition~\cite{REF:landau}. This Landau theory was constructed using
only symmetry considerations and the assumption that the phase
transition should be continuous (and thus dominated by the
long-distance behavior of the system).

By expanding any of the semi-microscopically derived free energy
functionals corresponding to the above-mentioned physical systems in
powers of the order parameter and gradients, one recovers this general
Landau free energy functional, up to terms that play no role in the
mean field theory, in the vicinity of the transition. In the study of
fluctuations, as we shall show later, these additional terms do not
alter the physical picture.  The slight difference between the general
Landau theory and the microscopically-derived theories is due to the
fact that, whereas the former theory only allows states that have a
disorder-averaged particle-density that is spatially homogeneous, in
the latter states with spatial inhomogeneities of the density are in
principle allowed, but are ultimately suppressed by the repulsive
interparticle interactions.

Let us begin by summarizing the results of the mean field theory of
the liquid--amorphous-solid transition driven by random constraints:
(i) for densities of constraints smaller than a critical value the
system is in the liquid state and all particles are delocalized; 
(ii) for densities of constraints larger than the critical value the
system is in an amorphous solid state, characterized by emerging
random static density-fluctuations; 
(iii) at the critical density of constraints there is a continuous
phase transition between the liquid and the amorphous solid states;
(iv) in the amorphous solid state a positive fraction of the particles
is localized around random mean positions and with random r.m.s.
displacements;
(v) in the amorphous solid state, close to the transition, the
fraction of localized particles is proportional to the excess
of the crosslink density beyond its critical value, and the typical
localization length diverges at the transition like the excess
crosslink density to the power $-1/2$;
(vi) when scaled by the mean value, the statistical distribution of
localization lengths is universal near the transition and,
consequently, the dependence of the order parameter on the wavevectors
also has a universal scaling form.

Although in the amorphous solid state translational invariance and
rotational invariance are broken at the microscopic level, because a
fraction of the monomers are localized, the average density is
uniform, and the system is macroscopically translationally (and
rotationally) invariant (MTI)~\cite{REF:cross}. The liquid state is,
of course, not only macroscopically but also microscopically
translationally and rotationally invariant, as each individual monomer
density is uniform over the container. Moreover, as in the case of
some of the mean field solutions encountered in spin glass systems,
both the liquid and the amorphous solid states are
replica-symmetric~\cite{REF:SK}.
However, as will be shown later, the symmetry of these states is even
larger, because the replica symmetry combines with the rotational
symmetry (in $d$-dimensional space) to produce rotational symmetry in a
$nd$-dimensional (\ie, replicated) space.

It is well known that in the spin glass case the early
replica-symmetric spin glass solution~\cite{REF:SK} for the
Sherrington-Kirkpatrick model was later found to be locally unstable
at temperatures below the critical temperature by de Almeida and
Thouless~\cite{REF:AT}, and was superseded by the
replica-symmetry-breaking solution~\cite{REF:Parisi} discovered by
Parisi, with its elegant interpretation in terms of symmetry-unrelated
pure equilibrium states~\cite{REF:Parisi_OP}. This
replica-symmetry-breaking solution has a nonnegative entropy at all
temperatures and is locally marginally
stable~\cite{REF:Parisi_stable}. 

By analogy with the spin glass case, and considering also the fact
that random topological constraints can, in principle, lead to a
partitioning of the configuration space of three-dimensional
macromolecular systems into ergodic regions that are not connected by
symmetry operations~\cite{REF:prl_1987,REF:cross}, it would not be
entirely surprising if in systems undergoing a liquid--amorphous-solid
transition under the effect of random constraints, the
replica-symmetric stationary point of the free energy corresponding to
the amorphous solid state turned out to be unstable and had to be
superseded by a less symmetric solution of the mean field equations.

However, the Deam-Edwards distribution~\cite{REF:DeamEd,REF:cross}
used to model the disorder favors sets of constraint locations that
are associated with highly probable configurations of the
unconstrained system. Thus, although frustration may in principle be
present, the disorder distribution tends to discourage
it~\cite{REF:Seung}. 

The purpose of this Paper is to show that the mean field amorphous
solid state is, in fact, locally stable, at least near the
transition. More specifically, the Hessian matrix that describes
changes of the free energy functional for infinitesimal fluctuations
around the stationary point corresponding to the amorphous solid state
is computed, and it is shown that, to linear order in the excess
crosslink density, all its eigenvalues are strictly positive, except
for a single zero mode associated with the spontaneous breaking of the
continuous translational symmetry of the system. In order to do this,
we construct a description for the space of fluctuations around the
stationary point that allows the eigenvalue problem for the Hessian
matrix in replicated space to be reduced, in essence, to an integral
eigenvalue equation in one dimension.

The rest of this Paper is organized as follows: In
Sec.~\ref{SEC:landau_matrix} we compute the Hessian matrix for the
Landau free energy functional around the amorphous solid state, and
make use of the continuous symmetry of the problem to identify a basis
set that significantly simplifies the eigenvalue equation for the
Hessian. In Sec.~\ref{SEC:landau_bound} we find positive lower bounds
for all the eigenvalues of the Hessian, except for the zero mode that
is present due to the spontaneously broken symmetry. In
Sec.~\ref{SEC:rcms} we extend these results to allow for fluctuations
in the particle density in the case of the microscopically-derived
free energy functional for randomly crosslinked
macromolecules. Finally, in Sec.~\ref{SEC:conclusions} we present our 
conclusions.

\section{Landau Theory: Hessian matrix}
\label{SEC:landau_matrix}

In this section we first summarize briefly the basics of the general
Landau theory of the liquid--amorphous-solid phase transition, as
presented in Ref.~\cite{REF:landau}, and later derive expressions for
the Hessian matrix for the free energy functional in this theory.

\subsection{Brief review of the Landau theory}
\label{SEC:landau_review}

In a system characterized by static random density fluctuations, the
appropriate order parameter
is~\cite{REF:prl_1987,REF:cross,REF:landau} 
\begin{equation}
\overline{\Omega}_{{\bf k}^{1},{\bf k}^{2},\cdots,{\bf k}^{g}}
\equiv
\left[
\frac{1}{N}\sum_{j=1}^{N}
\langle e^{i{\bf k}^{1}\cdot {\bf c}_{j}} \rangle_{\chi}
\langle e^{i{\bf k}^{2}\cdot {\bf c}_{j}} \rangle_{\chi}
\cdots
\langle e^{i{\bf k}^{g}\cdot {\bf c}_{j}} \rangle_{\chi}
\right], 
\label{EQ:opDefinition}
\end{equation}
where $N$ is the total number of particles, ${\bf c}_{i}$ (with
$i=1,\ldots, N$) is the ($d$-dimensional) position vector of particle
$i$, the wave-vectors ${\bf k}^{1},{\bf k}^{2},\cdots,{\bf k}^{g}$ are
arbitrary, $\langle\cdots\rangle_{\chi}$ denotes a thermal average for
a particular realization $\chi$ of the disorder, 
$\left[\cdots\right]$ represents averaging over the disorder, and $g$
is a positive number. 

We make the Deam-Edwards assumption~\cite{REF:DeamEd} that the
statistics of the disorder is determined by the correlations of the
unconstrained system.  Under the Deam-Edwards assumption, obtaining
disorder averages with the replica technique amounts to working with
the $n\to 0$ limit of systems of $n+1$, as opposed to $n$,
replicas. The additional replica, labeled by $\alpha=0$, represents
the degrees of freedom of the original system before adding the
constraints or, equivalently, describes the constraint distribution.

In the replica formalism, the order parameter takes the
form~\cite{REF:prl_1987,REF:cross,REF:landau} 
\begin{equation}
\overline{\Omega}_{\hat{k}}\equiv
\Big\langle
\frac{1}{N}\sum_{i=1}^{N}
\exp\big(i{\hat{k}}\cdot {\hat c}_{i}\big)
\Big\rangle_{n+1}^{\rm P}.
\label{EQ:ReplicaOrder}
\end{equation}
Here, hatted vectors denote replicated collections of
($d$-dimensional) vectors, viz., ${\hat{v}}\equiv ({\bf v}^{0},{\bf
v}^{1},\cdots,{\bf v}^{n})$, their scalar product being
${\hat{v}}\cdot {\hat{w}}\equiv \sum_{\alpha=0}^{n}{\bf v}^{\alpha}\cdot {\bf
w}^{\alpha}$, and $\langle\cdots\rangle_{n+1}^{\rm P}$ denotes an
average for an effective pure (\ie, disorder-free) system of $n+1$
coupled replicas of the original system. We use the terms {\it
one-replica sector\/} (1rs) and {\it higher-replica sector\/} (hrs) to
refer to replicated vectors with, respectively, exactly one and more
than one replica $\alpha$ for which the corresponding vector ${\bf
k}^{\alpha}$ is nonzero.

In the Landau theory the order parameter in the one-replica sector
represents spatial variations in the disorder-averaged mean
particle-density, and is always assumed to be strictly zero.

In the stationary-point approximation, the 
disorder-averaged free energy $f$ (per particle and space dimension) 
is given by ~\cite{REF:MPVbook,FNOTE:irrel_const,FNOTE:notation}
\begin{equation}
f=\lim_{n\rightarrow 0}\min_{\{\Omega_{\hat{k}}\}} 
{\cal F}_{n}\big(\{\Omega_{\hat{k}}\}\big),  
\label{EQ:physical_f}
\end{equation}  
with the Landau free energy functional given by
\begin{eqnarray}
&&nd{\cal F}_{n}\big(\{\Omega_{\hat{k}}\}\big) = 
{\overline{\sum}}_{\hat{k}}
\Big(-\epsilon+\frac{|\hat{k}|^2}{2}\Big)
\big\vert\Omega_{\hat{k}}\big\vert^{2}
\nonumber\\
&&\qquad\qquad
-\,{\overline{\sum}}_{{\hat{k}_1}{\hat{k}_2}{\hat{k}_3}}
\Omega_{\hat{k}_1}\,
\Omega_{\hat{k}_2}\,
\Omega_{\hat{k}_3}\,
\delta_{{\hat{k}_1}+{\hat{k}_2}+{\hat{k}_3}, {\hat{0}}}\,.
\label{EQ:LG_rescale}
\end{eqnarray}
Here $\epsilon$ is the control parameter, and is proportional to the
amount by which the constraint density exceeds its value at the
transition. The symbol ${\overline{\sum}}$ denotes a sum over
replicated wave vectors $\hat{k}$ in the higher-replica sector. 

For $\epsilon < 0$, the stationary-point equations for the free energy
functional of Eq.~(\ref{EQ:LG_rescale}) only have the solution
$\overline{\Omega}_{\hat{k}} = \delta_{\hat{k},\hat{0}}$, corresponding to the
liquid state. For $\epsilon > 0$, there are two solutions, one
($\overline{\Omega}_{\hat{k}} = \delta_{\hat{k},\hat{0}}$) 
corresponding to the liquid state and a second one corresponding
to an amorphous solid state, given by
\begin{eqnarray}
\overline{\Omega}_{\hat{k}}
&=&
\left(1-q\right)
\delta_{\hat{k},\hat{0}}+
q\,
\delta_{\tilde{\bf k},{\bf 0}}\,
\omega\left(\sqrt{2\hat{k}^{2}/\epsilon}\right),
\nonumber\\
q &=& 2\epsilon/3,
\nonumber\\
\omega(k)
&=&
\int_{0}^{\infty}d\theta\,\pi(\theta)
{\rm e}^{-k^{2}/2\theta}\,.
\label{EQ:ord_par_scale}
\end{eqnarray}
Here, the quantity $q$ is the ratio of the number of localized
particles to the total number of particles (\ie, the {\em localized
fraction}), $\pi(\theta)$ is a universal scaling function that
characterizes the distribution $p(1/\xi^{2})$ of the (inverse square)
localization lengths $\xi$ for the localized particles, through
$p(1/\xi^{2}) = (2/\epsilon) \pi(2/{\epsilon \xi^{2}})$, and we use
the definition $\tilde{\bf k} \equiv \sum_{\alpha=0}^{n} {\bf
k}^{\alpha}$. The factor $\delta_{\tilde{\bf k},{\bf 0}}$ encodes the
property of macroscopic translation invariance (MTI) for the amorphous
solid state, \ie, the fact that the state is invariant under common
translations of all the replicas, or, in more physical terms, that the
particles are localized around randomly located points that have a
homogeneous probability of being found anywhere in the volume of the
system. The scaling function $\pi(\theta)$ satisfies the stationarity
condition
\begin{equation}
\frac{\theta^{2}}{2} \frac{d\pi}{d\theta}
= (1-\theta)\,\pi(\theta)-
\int_{0}^{\theta} d\theta^{\prime}
\pi(\theta^{\prime})\pi(\theta-\theta^{\prime}), 
\label{EQ:scpieq}
\end{equation}
together with the normalization condition 
\begin{equation}
1=\int\nolimits_{0}^{\infty}d\theta\,\pi(\theta).
\label{EQ:pi_norm}
\end{equation}
This normalization condition directly follows from the fact that the
order parameter of Eq.~(\ref{EQ:ReplicaOrder}) has to be unity at the
origin of replicated wave vector space~\cite{REF:epl,REF:cross}.  It
is worth noticing that in the limit $\epsilon \rightarrow 0$ the above
parametrization of $\overline{\Omega}$ reduces continuously to the order
parameter $\overline{\Omega}_{\hat{k}} = \delta_{\hat{k},\hat{0}}$ for the liquid
state, as it should.

Let us now discuss the symmetry properties of the Landau free-energy
functional.  Under independent translations of all the replicas, \ie,
${\bf c}_{i}^{\alpha}\rightarrow{\bf c}_{i}^{\alpha}+{\bf
a}^{\alpha}$, the replica order parameter,
Eq.~(\ref{EQ:ReplicaOrder}), transforms as
\begin{equation}
\Omega_{\hat{k}}\rightarrow
\Omega^{\prime}_{\hat{k}}= 
{\rm e}^{i \hat{k}\cdot \hat{a}
} \, \Omega_{\hat{k}}.
\label{EQ:Omega_trans}
\end{equation}
For later reference, let us calculate the change in the order
parameter for the case of small displacements of the replicas:
\beq
\delta \Omega_{\hat{k}} \equiv \Omega^{\prime}_{\hat{k}} -
\Omega_{\hat{k}} = i \, \hat{k} \!\cdot \! \hat{a} \, \Omega_{\hat{k}}
+ {\cal O}(a^{2}).
\label{EQ:small_trans}
\eeq Under independent rotations of the replicas, defined by ${\bf
c}_{i}^{\alpha} \rightarrow {\bf c'}_{i}^{\alpha} = R^{\alpha}{\bf
c}_{i}^{\alpha}$, and ${\hat{R}}{\hat{v}}\equiv \{R^0{\bf
v}^{0},\cdots,R^{n}{\bf v}^{n}\}$, where each $R^{\alpha}$ is a
rotation matrix in $d$ dimensions, the order parameter transforms as
\begin{equation}
\Omega_{\hat{k}}\rightarrow
\Omega^{\prime}_{\hat{k}}=
\Omega_{\hat{R}^{-1}\hat{k}}.
\label{EQ:Omega_rot}
\end{equation}
By inserting the transformed order parameter for either of the above
operations into the free energy functional Eq.~(\ref{EQ:LG_rescale}),
we see that in both cases:
\beq
nd{\cal F}_{n}\big(\{\Omega'_{\hat{k}}\}\big) = 
nd{\cal F}_{n}\big(\{\Omega_{\hat{k}}\}\big), 
\label{EQ:F_symmetric}
\eeq
\ie, that the Landau free energy is invariant under {\em independent}
translations and rotations of the replicas. 

Anticipating the conclusions of this Paper, that the liquid state, 
which becomes unstable when $\epsilon$ is increased through zero, 
is replaced by a stable amorphous solid state for $\epsilon>0$, we 
now pause to compare some aspects of this phase transition with 
their counterparts in simple models of the paramagnet-to-ferromagnet 
phase transitions [e.g., the O($N$) symmetric vector $\phi^4$ model].
We shall refer to Fourier components of fields as {\it modes\/}, 
and shall consider the mean-field level of description.  At high 
temperatures (for magnetism) and low constraint densities (for 
amorphous solidification) the equilibrium value of all modes is zero. 
As the relevant control parameter is changed through its critical 
value, a band of modes, including those of the longest wave length, 
become linearly unstable, the longer the wave length the stronger the 
instability.  In both settings, magnetism and amorphous solidification, 
stability is recovered by the acquisition of a nonzero equilibrium 
value by one or more of the modes.  For magnetism, there is a {\it zero\/} 
wave vector mode, which is the most unstable mode, and by giving it the 
appropriate nonzero equilibrium value, this mode {\it and all others\/} 
are restabilized (i.e.~are no longer {\it un\/}stable---as discussed 
below, there should and does remain one marginally stable Goldstone mode).  
For amorphous solidification, restabilization is more intricate.  There is 
no fluctuating zero wave vector mode in the theory to be given a nonzero 
equilibrium value.  Instead, the most unstable modes have the smallest 
allowed nonzero wave vectors.  If these modes become nonzero, as some of 
them do, there is no symmetry-dictated \lq\lq selection rule\rq\rq\ 
prohibiting them from acting as \lq\lq sources\rq\rq\ for certain other 
modes, and thus not just one but a large family of modes become nonzero, not 
only including modes that were formerly unstable.  (This may be regarded as 
an analog of domain wall formation in wave vector space.)\thinspace\ There 
is one further subtlety to the amorphous solidification case: to give a 
nonzero equilibrium value to modes that reside in the one replica sector 
would be extremely energetically costly and, from the viewpoint of the 
Landau theory, is ruled out (via an implicit linear constraint on the 
order parameter).  This requirement is satisfied by giving a nonzero 
equilibrium value only to modes that are MTI, because such modes are 
prohibited, on symmetry grounds, from acting as sources for modes in 
the one replica sector.  Thus, stability is restored not by giving one 
unstable mode a nonzero value, and not by giving only the unstable modes a 
nonzero value, but by giving a sheet of modes (some unstable and some 
stable) a nonzero value.

\subsection{Hessian matrix elements}
\label{SEC:H_pq}

Consider any variation $\{\delta\Omega_{\hat{k}}\}$ of the
$\{\Omega_{\hat{k}}\}$ around a stationary point
$\{\overline{\Omega}_{\hat{k}}\}$ . To first order in
$\{\delta\Omega_{\hat{k}}\}$, the variation of the free energy
functional is, of course, zero. We see from Eq.~(\ref{EQ:LG_rescale})
that the second order variation is~\cite{FNOTE:support}
\bea
&&{\delta}^{(2)}{\left[nd{\cal F}_{n}\big(\{\Omega_{\hat{k}}\}\big)\right]} = 
{\overline{\sum}}_{\hat{k}}
\left(-\epsilon+\frac{|\hat{k}|^2}{2}\right)
\big\vert \delta \Omega_{\hat{k}} \big\vert^{2}
\nonumber\\
&& \quad
-3\,{\overline{\sum}}_{{\hat{k}_1}{\hat{k}_2}{\hat{k}_3}}
\delta_{{\hat{k}_1}+{\hat{k}_2}+{\hat{k}_3}, {\hat{0}}}\,
\overline{\Omega}_{\hat{k}_1}\,
\delta\Omega_{\hat{k}_2}\,
\delta\Omega_{\hat{k}_3}\,.
\label{EQ:F2_variation}
\eea
Now consider expanding around the liquid state for any value of
$\epsilon$. In this case, the second variation reduces to
\beq
\delta^{(2)} {\left[nd{\cal F}_{n}\big(\{\Omega_{\hat{k}}\}\big)\right]} = 
{\overline{\sum}}_{\hat{k}}
\left(-\epsilon+\frac{|\hat{k}|^2}{2}\right)
\big\vert \delta \Omega_{\hat{k}} \big\vert^{2},
\label{EQ:F2_liquid}
\eeq which evidently indicates that the liquid is stable for $\epsilon
< 0$ and unstable for $\epsilon > 0$.  For $\epsilon > 0$ the only
candidate we know of for a stable thermodynamic state is the amorphous
solid. {}From now on we focus only on that state.

By inserting the value of the order parameter,
Eq.~(\ref{EQ:ord_par_scale}), into the three-wavevector sum in
Eq.~(\ref{EQ:F2_variation}), we obtain 
\bea
&& {\overline{\sum}}_{{\hat{k}_1}{\hat{k}_2}{\hat{k}_3}}
\delta_{{\hat{k}_1}+{\hat{k}_2}+{\hat{k}_3}, {\hat{0}}}\,
\, \delta_{{\tilde{\bf k}_1}, {\bf 0}} \, \frac{2 \epsilon}{3} 
\, \omega\!\left(\sqrt{2\hat{k}_1^{2}/\epsilon}\right)
\delta\Omega_{\hat{k}_2}\,
\delta\Omega_{\hat{k}_3}\,
\nonumber \\ 
&& = {\overline{\sum}}_{{\hat{k}_2}{\hat{k}_3}}
\delta_{{\tilde{\bf k}_2}+{\tilde{\bf k}_3}, {\bf 0}} \, \frac{2
\epsilon}{3}  
\, \omega\!\left(\sqrt{\frac{2}{\epsilon} \left(\hat{k}_2+\hat{k}_3
\right)^{2}}\right) 
\delta\Omega_{\hat{k}_2}\,
\delta\Omega_{\hat{k}_3}\, 
\nonumber \\ 
&& \qquad - \frac{2 \epsilon}{3} {\overline{\sum}}_{\hat{k}}
\big\vert \delta \Omega_{\hat{k}} \big\vert^{2}.
\label{EQ:sum_arrange}
\eea
Thus we can rewrite the second variation in terms of
the Hessian matrix $H_{\hat{q}, \hat{q}'}$: 
\begin{equation}
\delta^{(2)} \left[nd{\cal F}_{n}(\{\Omega_{\hat{q}}\}) \right] = 
\overline{\sum}_{\hat{q}, \hat{q}'} H_{\hat{q}, \hat{q}'} \,
\delta \Omega_{\hat{q}} \, \delta \Omega_{-\hat{q}'} \, ,
\label{EQ:Taylor}
\end{equation}
where we have defined $H_{\hat{k}, \hat{l}}$ by
\beq
H_{\hat{k}, \hat{l}} 
\equiv  \frac{1}{2!} \frac{ \delta^{2} \left[ nd{\cal F}_{n} \right] }  
{\delta \Omega_{\hat{k}} \delta \Omega_{-\hat{l}}}.
\label{EQ:hessian_def}
\eeq
(For later convenience, we have chosen a definition that differs from
the standard one by a factor of $1/2$.) 
More explicitly, we have  
\bea
H_{\hat{k}, \hat{l}} 
& = & \delta_{{\hat{k}},{\hat{l}}} 
\left(\epsilon + \frac{{\hat{k}}^{2}}{2} \right) 
- \delta_{\tilde{\bf k},\tilde{\bf l}} \, 2 \epsilon 
\int_{0}^{\infty} \!\! d\theta \, \pi(\theta) \, {\rm
e}^{-({\hat{k}}-{\hat{l}})^{2}/{\epsilon \theta}}
\nonumber \\
&&
\qquad \qquad \qquad + {\cal O}(\epsilon^{2}). 
\label{EQ:F2_hrs}
\eea
As it is adequate to be concerned with matrix elements to leading
(\ie~first) order in $\epsilon$, we shall neglect higher orders from
now on.

\subsection{Change of basis}
\label{SEC:H_change}

In order to simplify the diagonalization of the Hessian, we are going
to exploit the symmetries of the problem.  As a direct consequence of
the invariance of the free-energy functional under translations and
the MTI property of the amorphous solid state, the matrix element
$H_{\hat{k}, \hat{l}}$ of the Hessian only connects wavevectors
$\hat{k}$ and $\hat{l}$ such that $\tilde{\bf k} = \tilde{\bf
l}$. This already reduces the complexity of the problem by making the
Hessian block diagonal.

As $H_{\hat{k}, \hat{l}}$ depends on ${\hat{k}}^{2}$, ${\hat{l}}^{2}$,
and ${\hat{k}}\cdot {\hat{l}}$, one might expect to find a symmetry under
arbitrary rotations in $(n+1)d$ dimensions, which would simplify the
diagonalization of the Hessian still further. However, the factor
$\delta_{\tilde{\bf k},\tilde{\bf l}}$ is not invariant under some of
those rotations. Instead, $H$ only displays a rotational symmetry in
$nd$ dimensions, but this will enable us to simplify the task in
much the same way as is commonly done for central potentials in
quantum mechanics.

In order to make this symmetry explicit, we choose a {\em fixed}
matrix $T \in SO((1+n)d)$ such that for any vector $\hat{v}$ in
replicated space we explicitly isolate ${\bf \tilde{v}}$ from the
other $nd$ independent coordinates, which we call $\breve{v}$:
\beq
T \hat{v} = \left( \begin{array}{c}
                         \frac{{\bf \tilde{v}}}{\sqrt{1+n}} \\
                         \breve{v}
                   \end{array}      \right).
\label{EQ:T_def}
\eeq
Due to $T$ being orthogonal, scalar products remain simple in the 
new coordinates:
\beq
\hat{v} \cdot \hat{w} = \frac{{\bf \tilde{v}}\cdot {\bf \tilde{w}}}{1+n} + 
\breve{v} \cdot \breve{w}. 
\label{EQ:scalar_product}
\eeq
In the new coordinates, we have 
\bea
H_{\tilde{\bf k} \breve{k}, \, \tilde{\bf l} \breve{l}} 
& = & \delta_{\tilde{\bf k},\tilde{\bf l}} 
\Big\{ 
\, \delta_{{\breve{k}},{\breve{l}}} 
\Big[ \, \epsilon + \frac{1}{2} \big({\breve{k}}^{2} + \frac{{\tilde{\bf
k}}^{2}}{1+n} \big) \Big] \Big.
\nonumber \\
&&
\quad \Big. - 2 \epsilon 
\int_{0}^{\infty} \!\! d\theta \, \pi(\theta) \, {\rm
e}^{-({\breve{k}}-{\breve{l}})^{2}/{\epsilon \theta}}
\Big\}.
\label{EQ:F2_T}
\eea
This expression, taken na\"{\i}vely, would immediately tell us that
the Hessian is invariant under rotations in $nd$ dimensions:
\beq
\forall \breve{R} \in SO(nd): \qquad \qquad 
H_{\tilde{\bf k} \, \breve{R}\breve{k}, \,\, \tilde{\bf l} \,
\breve{R}\breve{l}}  
= H_{\tilde{\bf k} \breve{k}, \, \tilde{\bf l} \breve{l}}. 
\label{EQ:F2_rot}
\eeq 
However, there is an important caveat. Our Hessian is only
defined for wavevectors in the higher replica sector, but the proposed
rotations can take a vector in the higher replica sector and transform
it into a vector in the one replica sector. For the moment we are
going to ignore this difficulty, and simply diagonalize the matrix
obtained by using Eq.~(\ref{EQ:F2_hrs}) as its definition, with
$\hat{k}$ and $\hat{l}$ taking {\em any} nonzero values, both in the
higher and in the one replica sector. (This enlarged Hessian matrix
will be termed the ``extended Hessian'' to distinguish it from the
``original Hessian'' which does not have those additional matrix
elements. See Fig.~\ref{FIG:extended_H}.) After having diagonalized
the extended Hessian, we will 
return to the issue of the one replica sector. For the moment, let us
just anticipate that, in the replica limit $n \to 0$, the only effect
of this extension of the Hessian on its spectrum of eigenvalues will
be the addition of one spurious eigenvalue, corresponding to a
fluctuation localized in the one replica sector.

%
\begin{figure}[htbp]
\centerline{\psfig{figure=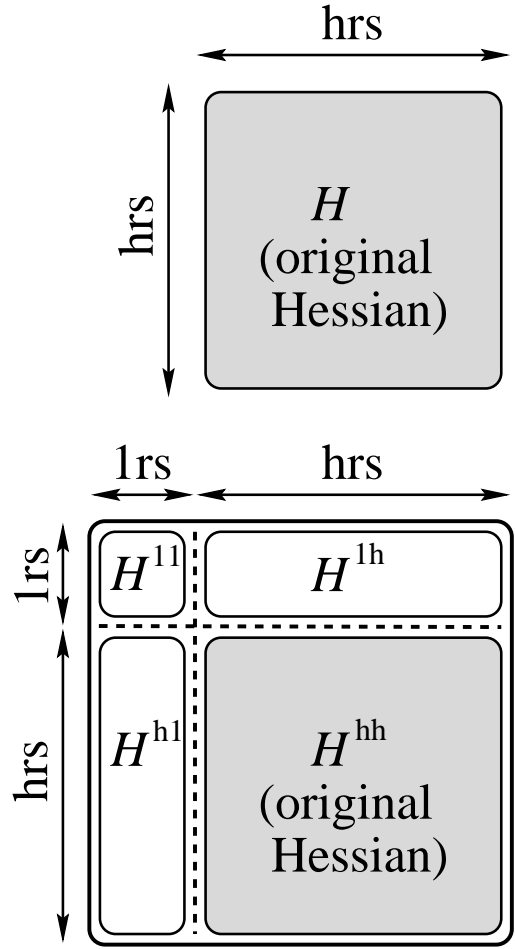,height=12.5cm}}
\nopagebreak
\vspace{0.7cm}
\nopagebreak
\narrowtext 
\centerline{\caption{
Comparison of the original Hessian matrix (upper figure) and the
extended Hessian matrix (lower figure). The extended
Hessian matrix is depicted as being formed by four blocks, respectively
connecting: the 1rs with itself (labeled $H^{\rm 11}$), the 1rs
with the hrs (labeled $H^{\rm 1h}$), the hrs with the 1rs
(labeled $H^{\rm h1}$), and the hrs with the hrs (labeled
$H^{\rm hh}$). The block $H^{\rm hh}$ corresponds exactly to the
original Hessian. \label{FIG:extended_H}} 
}
\end{figure}
%

For the modified problem of diagonalizing the extended Hessian, the
$SO(nd)$ symmetry holds, and Eq.~(\ref{EQ:F2_rot}) is correct without
caveat. In what follows, we will use the same strategies as in the
diagonalization of a quantum-mechanical hamiltonian for a particle in
a central potential.  The role of the hamiltonian will be played by
$H_{\hat{k},\hat{l}}$.  We will also exploit the symmetries of the
problem: the rotational symmetry in $nd$ dimensional space will allow
us to write each eigenfunction as a product of a radial part, which
will be obtained by solving a one dimensional eigenvalue equation, and
an angular part, which will simply be a surface harmonic
function~\cite{REF:harmonics} in $nd$ dimensions. The quantity ${\bf
\tilde{k}}$, which is exactly conserved by $H_{\hat{k},\hat{l}}$, will
play the role of a conserved quantum number.  We will obtain distinct
sets of eigenfunctions for each fixed value of ${\bf \tilde{k}}$.

We are going to work in the Hilbert space of complex functions of the
variable $\hat{k}$ ($\neq \hat{0}$). We exclude the origin because we
are interested in fluctuations of the order parameter, and
$\Omega_{\hat{0}}$ cannot fluctuate. We define the scalar product in
this space as follows:
\beq
\langle f | g \rangle \equiv \frac{1}{V^{n}} 
\sum_{\hat{k} \neq \hat{0}} f^{\ast}(\hat{k}) \, g(\hat{k}),
\label{EQ:scalar_product_def}
\eeq
which simplifies, in the thermodynamic limit, to~\cite{FNOTE:remove_origin}
\bea
\langle f | g \rangle & \simeq & V \int \frac{d{\hat{k}}}{(2 \pi)^{(1+n)d}}
f^{\ast}(\hat{k}) \, g(\hat{k}) \nonumber \\
& = & V \! \int \! \frac{d{\bf \tilde{k}} \, d{\breve{k}}}
{(1+n)^{d/2} \, (2 \pi)^{(1+n)d}}
f^{\ast}({\bf \tilde{k}},{\breve{k}}) \, g({\bf \tilde{k}},{\breve{k}}).
\label{EQ:scalar_product_thermo}
\eea
We define a basis set for this Hilbert space by
\bea
\lefteqn{ \varphi_{p {\bf \tilde{p}} \sigma} ({\bf \tilde{k}},
\breve{k}) \equiv } \nonumber \\
&& \quad (1\!+\!n)^{d/4} \, (2 \pi)^{nd/2} \, 
\delta_{{\bf \tilde{p}}, {\bf \tilde{k}}} \,
\delta(|\breve{k}|\!-\!p) \, p^{\frac{1-nd}{2}} \,
S_{\sigma}(\phi_{\breve{k}}),
\label{EQ:basis}
\eea 
Here $\{S_{\sigma}(\phi) \}$ are the normalized surface harmonic
functions defined on the unit sphere in $nd$-dimensional
space~\cite{REF:harmonics}, ${\bf \tilde{p}}$ is any wavevector in $d$
dimensions, and $p$ is any positive number. (Surface harmonics are
homogeneous trigonometric polynomials, they are generalizations of
spherical harmonics to any space dimension $d \geq 3$. The label
$\sigma$ is a set of integers that characterize the appropriate
trigonometric polynomial: for example, for $d=3$, the surface
harmonics are the usual spherical harmonics, and $\sigma \equiv
(l,m)$, with $l$ the degree of the trigonometric polynomial and $m$ a
label that distinguishes between polynomials of the same degree.) The
notation $\phi_{\breve{k}} \equiv {\breve{k}}/|{\breve{k}}|$ denotes
the unit $nd$-dimensional vector along the direction of
${\breve{k}}$. The elements of the basis set $\{ \varphi_{p {\bf
\tilde{p}} \sigma} \}$ are orthogonal and normalized under the scalar
product Eq.~(\ref{EQ:scalar_product_def}):
\beq 
\langle \varphi_{p' {\bf \tilde{p}'} \sigma'} 
        | \varphi_{p {\bf \tilde{p}} \sigma} \rangle =
\delta(p' - p) \, \delta_{ {\bf \tilde{p}'}, {\bf \tilde{p}} } \,
\delta_{\sigma', \sigma}.
\label{EQ:orthonormality}
\eeq

As suggested above, we propose to express each eigenfunction for the
problem in the form:
\bea
\lefteqn{ \psi_{r {\bf \tilde{p}} \sigma} ({\bf
\tilde{k}},{\breve{k}}) }
\nonumber \\
&& \qquad = 
(1\!+\!n)^{d/4} (2 \pi)^{nd/2} \, 
\delta_{{\bf \tilde{p}}, {\bf \tilde{k}}} \, 
|\breve{k}|^{\frac{1-nd}{2}} \, R_{r}(|{\breve{k}}|) \, 
S_{\sigma}(\phi_{\breve{k}}),
\nonumber \\
&& \qquad = 
\int_{0}^{\infty} \!\! dp \, R_{r}(p) \, 
\varphi_{p {\bf \tilde{p}} \sigma} ({\bf \tilde{k}},{\breve{k}}),
\label{EQ:eigenfunction_prod}
\eea 
where the discrete $\delta$ function ensures that the eigenfunction 
is localized on points with a fixed value of ${\bf \tilde{k}}$, 
the surface harmonic $S_{\sigma}$ gives the angular dependence on
${\breve{k}}$, and the radial function $R_{r}$ gives the radial 
dependence on ${\breve{k}}$. By using the normalization condition for the 
basis set, we obtain the normalization condition for the radial part:
\beq
\int_{0}^{\infty} \! dk \, |R_{r}(k)|^{2} = 1
\label{EQ:R_norm}
\eeq
We now compute the matrix elements of the Hessian between the elements
of the basis set $\{\varphi_{p {\bf \tilde{p}} \sigma}\}$. 

Consider first the part $H^{D}$ of the Hessian that is diagonal in
$\breve{k}$, \ie, the first term on the \rhs of
Eq.~(\ref{EQ:F2_T}). This part is also diagonal in the basis
$\{\varphi_{p {\bf \tilde{p}} \sigma}\}$, with the matrix elements
\bea
\lefteqn{ \langle \varphi_{p' {\bf \tilde{p}'} \sigma'} | H^{D}
| \varphi_{p {\bf \tilde{p}} \sigma} \rangle = } \nonumber \\
&& \delta(p' - p) \, \delta_{ {\bf \tilde{p}'}, {\bf \tilde{p}} } \,
\delta_{\sigma', \sigma} \,
\Big[ \, \epsilon + \frac{1}{2} \Big({\breve{p}}^{2} + \frac{{\tilde{\bf
p}}^{2}}{1+n} \Big) \, \Big] .
\label{EQ:HD}
\eea 

The non-diagonal part $H^{O}$ of the Hessian, given by the second term
on the \rhs of Eq.~(\ref{EQ:F2_T}), has, in the new basis, matrix
elements that only connect different values of the radial coordinate
$p$ but are still diagonal in ${\bf \tilde{p}}$ and $\sigma$. It is
shown in App.~\ref{APP:matrix_el} that the matrix elements of $H^{O}$
have the form
\bea
\lefteqn{ \langle \varphi_{p' {\bf \tilde{p}'} \sigma'} | H^{O}
| \varphi_{p {\bf \tilde{p}} \sigma} \rangle = } \nonumber \\
&&
\quad
\delta_{{\bf \tilde{p}'}, {\bf \tilde{p}}}  \,
\delta_{\sigma', \sigma} \, 
(-2 \epsilon) \, \epsilon^{-1/2} \, C_{n} 
\, \eta^{(n)}_{|\sigma|}(p'/\sqrt{\epsilon},p/\sqrt{\epsilon}),
\label{EQ:HO_res}
\eea 
with 
\beq
C_{n} \equiv \left(\epsilon/{4\pi}\right)^{nd/2} (1+n)^{-d/2} ,
\label{EQ:cn_def}
\eeq 
and 
\bea
\lefteqn{ \eta^{(n)}_{l} (x',x) \equiv 2 \sqrt{x' x} }
\nonumber \\ 
&& \quad \times 
\int\limits_{0}^{\infty} \! \frac{d\theta \,
\pi(\theta)}{\theta^{1-nd/2}} \,  
{\rm e}^{-(x'^2 + x^2)/{\theta}}  
\, I_{l-1+nd/2}\left(\frac{2 x' x}{\theta}\right), 
\label{EQ:eta_def}
\eea
where $I_{\nu}(x)$ is the modified Bessel function of order
$\nu$. The label $l (\equiv |\sigma|) $ indicates the degree of the
surface harmonic $S_{\sigma}$ as a trigonometric polynomial.  The
constant $C_{n}$ satisfies the condition 
\beq 
\lim_{n \to 0} C_{n} = 1,
\label{EQ:cn_lim}
\eeq
via which it disappears from the eigenvalue equation in the replica
limit.  The kernel $\eta^{(n)}_{l} (x',x)$ is real and symmetric,
and controls the non-diagonal nature of the matrix elements. Due to the
positivity of $I_{\nu}(y)$ for $\nu \geq -1$ and $y > 0$, the kernel 
$\eta^{(n)}_{l} (x',x)$ is positive for $x x' > 0$. 
For $x x' \to 0^{+}$, $\eta^{(n)}_{l} (x',x)$ vanishes, except if
$l=0$ and $nd > 0$, in which case it is divergent.

As both $H^{D}$ and $H^{O}$ are diagonal on the ${\bf \tilde{p}}$
and $\sigma$ labels, the eigenvalue equation for the Hessian,
\beq
H | \psi \rangle = \kappa | \psi \rangle,
\label{EQ:eigen_def} 
\eeq
can now be simplified to a radial equation:
\bea 
\lefteqn{ \kappa R(p) } \nonumber \\
& = & \int_{0}^{\infty} dp' \, 
\langle \varphi_{p {\bf \tilde{p}} \sigma} | H
| \varphi_{p' {\bf \tilde{p}} \sigma} \rangle 
R(p'),
\nonumber \\ 
& = & 
\Big[ \, \epsilon + \frac{1}{2} \big(p^{2} + \frac{{\tilde{\bf
p}}^{2}}{1+n} \Big) \, \Big] \, R(p) 
\nonumber \\
&& - 2 \, \epsilon \, C_{n} \int_{0}^{\infty} \frac{dp'}{\sqrt{\epsilon}} \,
\eta^{(n)}_{|\sigma|}(p/\sqrt{\epsilon},p'/\sqrt{\epsilon}) 
\,\,R(p').
\label{EQ:eigen_rad}
\eea
This radial equation can be simplified further by the rescaling
\bml
\bea
\zeta & = & \frac{1}{\epsilon} \Big[ \kappa- \Big(\epsilon + \frac{{\tilde{\bf
p}}^{2}}{2(1+n)} \Big) \, \Big] \, , 
\\
x & = & p/\sqrt{\epsilon}, \qquad
u(x) = \epsilon^{1/4} R(\sqrt{\epsilon} x),
\label{EQ:radial_scale}
\eea
\eml
which removes the $\epsilon$ dependence from the eigenvalue equation,
thus making both the eigenvalue $\zeta$ and the
eigenfunction $u(x)$ $\epsilon$-independent and obeying
\beq
\zeta u(x) = \frac{x^{2}}{2} u(x) -2 \, C_{n} \int_{0}^{\infty} \!\!
dx' \,\eta^{(n)}_{|\sigma|}(x,x') \,\, u(x').
\label{EQ:eigen_scaled_n}
\eeq
In the replica limit, $n \to 0$, this radial equation reduces to
\beq
\zeta u(x) = \frac{x^{2}}{2} u(x) -2 \int_{0}^{\infty} \!\!
dx' \,\eta^{(0)}_{|\sigma|}(x,x') \,\, u(x').
\label{EQ:eigen_scaled_0}
\eeq

For all cases except $|\sigma| = 0$ this limit is straightforward, because
$\eta^{(n)}_{|\sigma|}(x,x')$ smoothly converges to
$\eta^{(0)}_{|\sigma|}(x,x')$.
For the special case of $|\sigma| = 0$, the limit $n \to 0$ for
$\eta^{(n)}_{|\sigma|}(x,x')$ is singular near the origin.
Let us mention here a property of Eq.~(\ref{EQ:eigen_scaled_0}) that
does {\em not} apply to Eq.~(\ref{EQ:eigen_scaled_n}). As $I_{-1}(z) =
I_{1}(z)$ for all values of the variable $z$, we have the equality
\beq 
\eta^{(0)}_{0}(x,x') = \eta^{(0)}_{2}(x,x'),
\label{EQ:0_2}
\eeq 
which means that the radial equation~(\ref{EQ:eigen_scaled_0}) is
the same for $|\sigma| = 0$ and for $|\sigma| = 2$. 
In Sec.~\ref{SEC:1rs} we discuss in more detail the relations between
the solutions to Eq.~(\ref{EQ:eigen_scaled_n}) and
Eq.~(\ref{EQ:eigen_scaled_0}). 

Both radial equations, Eq.~(\ref{EQ:eigen_scaled_n}) and
Eq.~(\ref{EQ:eigen_scaled_0}), are eigenvalue equations for Hermitian
operators. This guarantees the existence of a complete orthonormal
basis of eigenfunctions, all of them having real eigenvalues. Notice
also the nontrivial fact that the radial equation is well defined in
the replica limit $n \to 0$.

The form of the radial eigenvalue equation tells us that the radial
eigenfunction and the eigenvalue depend on the degree $l = |\sigma|$
of the surface harmonic considered (which plays a role analogous to
that of the total angular momentum quantum number $l$ in the central
potential problem for a quantum mechanical particle), and on an
additional label $r$, playing a role analogous to the radial quantum
number in quantum mechanics. Therefore, the eigenvalues of the
extended Hessian are given by the relation: 
\beq
\kappa_{lr}(\tilde{\bf k}) = (1 + \zeta_{lr}) \epsilon +
\frac{{\tilde{\bf k}}^{2}}{2}.
\label{EQ:eig_gen}
\eeq

As it is easier to work with scaled variables, let us express the
condition that there be no unstable fluctuation directions (\ie,
$\kappa \geq 0$) in terms of $\zeta$: 
\beq
\kappa_{lr}(\tilde{\bf k}) > 0  \qquad \forall \tilde{\bf k} \iff \zeta_{lr}
+ 1 > 0. 
\label{EQ:cond_zeta}
\eeq
The right hand side of the equivalence sign is the condition that we
are going to establish in what follows.

\section{Landau Theory: eigenvalues of the Hessian}
\label{SEC:landau_bound}
In this section we are going to establish positive lower bounds for
all the eigenvalues of the Hessian of the Landau theory, except for
the zero mode associated with the spontaneously broken translational
symmetry. In other words, we study the set ${\cal S}_{o}$ containing
the limits, when $n \to 0$, of the eigenvalues of the {\em original\/}
Hessian.  However, for technical reasons, it is convenient to first
study two other sets of numbers, denoted by ${\cal S}_{e}$ and ${\cal
S}_{r}$. ${\cal S}_{e}$ is the set containing the $n \to 0$ limits of
the eigenvalues of the {\em extended} Hessian; each element in ${\cal
S}_{e}$ can be written in the form given by Eq.~(\ref{EQ:eig_gen}),
where $\zeta_{lr}$ is taken to be the $n \to 0$ limit of an eigenvalue
in Eq.~(\ref{EQ:eigen_scaled_n}). ${\cal S}_{r}$ is the set containing
all numbers $\kappa_{lr}(\tilde{\bf k})$ computed according to
Eq.~(\ref{EQ:eig_gen}), with $\zeta_{lr}$ chosen to be an eigenvalue
in the {\em radial} equation~(\ref{EQ:eigen_scaled_0}).

In Sec.~\ref{SEC:zero_mode} we show that ${\cal S}_{r}$ contains a
zero element corresponding to the zero mode associated with the
spontaneously broken translational symmetry. In
Sec.~\ref{SEC:lower_bounds} we compute positive lower bounds for all
other elements of ${\cal S}_{r}$. In Sec.~\ref{SEC:1rs}, we show that
the only difference between the limit for $n \to 0$ of the eigenvalue
spectrum of Eq.~(\ref{EQ:eigen_scaled_n}) and the eigenvalue spectrum
of Eq.~(\ref{EQ:eigen_scaled_0}) is that in the former a spurious
eigenvalue corresponding to fluctuations in the 1rs appears, which is
not present in the latter. Therefore the spectrum ${\cal S}_{e}$ of
the extended Hessian contains a spurious eigenvalue not present in
${\cal S}_{r}$. We also show, in Sec.~\ref{SEC:1rs}, that the
eigenvectors of the original Hessian correspond to all the
eigenvectors of the extended Hessian, except the spurious one, \ie~that
${\cal S}_{o}$ and ${\cal S}_{r}$ are identical. Using the results
obtained in Secs.~\ref{SEC:zero_mode} and \ref{SEC:lower_bounds}, this
will allow us to conclude that the amorphous solid state is locally
stable.

\subsection{Obtaining the zero mode}
\label{SEC:zero_mode}

We first consider the eigenfluctuation associated with the
translational symmetry, and show that it is a zero mode. 
{}From Eq.~(\ref{EQ:small_trans}), we see that this fluctuation can be
written as 
\bea
\delta \Omega_{\hat{k}} & = & i \hat{k}\!\cdot \!\hat{a} \, ({2
\epsilon}/{3}) \,
\delta_{\tilde{\bf k}, {\bf 0}}  
\int_{0}^{\infty} \!\! d\theta\,\pi(\theta)
\,{\rm e}^{-\hat{k}^{2}/\epsilon\theta}\,
\nonumber \\
& = & i \breve{k}\!\cdot \!\breve{a} \, ({2
\epsilon}/{3}) \,
\delta_{\tilde{\bf k}, {\bf 0}} 
\int_{0}^{\infty} \!\! d\theta\,\pi(\theta)
\,{\rm e}^{-\breve{k}^{2}/\epsilon\theta}\,.
\label{EQ:zero_fluct}
\eea
The only angular dependence of $\delta \Omega_{\hat{k}}$ is given by
the prefactor $\breve{k}\!\cdot \!\breve{a}$, which is a degree-one 
polynomial in $\breve{k}$. This guarantees that this fluctuation
resides in the $|\sigma| = 1$ sector. By taking
the scalar product with the appropriate element in the basis
$\{\varphi_{p {\bf \tilde{p}} \sigma}\}$ [which we label by $\sigma =
(1,0)$ by analogy to the spherical harmonic $Y_{10} \propto z/r$], we
obtain the radial function associated with $\delta \Omega_{\hat{k}}$:
\bea
R(k) & = & \langle \varphi_{p,{\bf \tilde{p}}={\bf 0},\sigma = (1,0)} |
\delta \Omega \rangle, 
\nonumber \\
& = & i A_{n} \, \epsilon \, k^{(1+nd)/2} \int_{0}^{\infty} \!\!
d\theta\,\pi(\theta) \,{\rm e}^{-k^{2}/\epsilon\theta}\,,
\label{EQ:radial_fluct}
\eea 
where $A_{n}$ is a numerical prefactor, which we can ignore in what
follows. Taking the replica limit, and transforming to scaled
variables, we obtain the scaled radial function
\beq
u(x) = \sqrt{x} \int_{0}^{\infty} \!\!
d\theta\,\pi(\theta) \,{\rm e}^{-x^{2}/\theta}\,.
\label{EQ:radial_fluct_scaled}
\eeq
In App.~(\ref{APP:radial_zero}) we show by explicit computation that
this form for $u(x)$ satisfies Eq.~(\ref{EQ:eigen_scaled_0}) with
$\zeta = -1$. By Eq.~(\ref{EQ:eig_gen}), this  means that the
corresponding $\delta \Omega_{\hat{k}}$ given by
Eq.~(\ref{EQ:zero_fluct}) is an eigenvector of the Hessian with zero
eigenvalue. 
[As $\delta \Omega_{\hat{k}}$ given by Eq.~(\ref{EQ:zero_fluct}) is
only nonzero for $\hat{k}$ in the hrs because otherwise $\breve{k} =
\breve{0}$, it is simultaneously an 
eigenvector of the extended Hessian and of the original Hessian.]

\subsection{Positive lower bounds for the eigenvalues}
\label{SEC:lower_bounds}

Having obtained the zero mode for a specific form of fluctuation, we
now discuss generic fluctuations in the order parameter field, and
show that all the other eigenvalues are positive-definite. For the
case $l \neq 1$, we will obtain positive lower bounds for the
eigenvalues by analytical manipulation of
Eq.~(\ref{EQ:eigen_scaled_0}). For the case $l = 1$, we will solve
Eq.~(\ref{EQ:eigen_scaled_0}) numerically and show explicitly that the
lowest eigenvalue corresponds to the zero mode already obtained, and
that all other eigenvalues correspond to positive values of
$\kappa(\tilde{\bf k})$.

Consider one particular scaled radial eigenfunction $u(x)$ in
Eq.~(\ref{EQ:eigen_scaled_0}), with eigenvalue $\zeta$.
To simplify the argument we temporarily switch to the normalization 
\beq
\int_{0}^{\infty} \!\! dx \,|u(x)| = 1,
\label{EQ:radial_norm}
\eeq
and we define the quantity 
\beq
s(x) \equiv {\rm sgn}\big(u(x)\big).
\label{EQ:sign_def}
\eeq
In what follows, we express the eigenvalue $\zeta$ in an unusual but
convenient form that allows a lower bound to be derived from
it. By combining the eigenvalue equation~(\ref{EQ:eigen_scaled_0}),
with the normalization condition (\ref{EQ:radial_norm}) and the
definition (\ref{EQ:sign_def}), we obtain
\bea
\zeta & = & \zeta \int_{0}^{\infty}\!\! dx \,|u(x)|
\nonumber \\
& = & \int_{0}^{\infty}\!\! dx \,s(x) \,\zeta \,u(x)
\nonumber \\
& = & \int_{0}^{\infty}\!\! dx \,s(x) \Big[ \frac{x^{2}}{2} u(x) 
-2 \!\int_{0}^{\infty}\!\! dx' \,\eta^{(0)}_{|\sigma|}(x,x') \,u(x')
\Big] \nonumber \\
& = & \int_{0}^{\infty}\!\! dx  \,\frac{x^{2}}{2} \,|u(x)| 
\nonumber \\
&& \qquad 
-2 \!\int_{0}^{\infty}\!\! dx dx' \,\eta^{(0)}_{|\sigma|}(x',x) 
\,|u(x)| \,s(x) \,s(x'). 
\label{EQ:bound_equal}
\eea
(In the last line we have interchanged the dummy variables $x$ and $x'$.)
{}From the expression just derived for the eigenvalue $\zeta$, it
follows from the nonnegativity of $\eta^{(0)}_{|\sigma|}(x',x)$ that 
\bml
\bea
\zeta & \geq & \int_{0}^{\infty}\!\! dx \,\gamma_{|\sigma|}(x) \,|u(x)| 
\geq \bar{\gamma}_{|\sigma|},
\label{EQ:bound_bound}
\\
\gamma_{l}(x) & \equiv & \frac{x^{2}}{2} 
-2 \!\int_{0}^{\infty}\!\! dx' \,\eta^{(0)}_{l}(x',x),
\\
\bar{\gamma}_{l} & \equiv &  \inf_{x} \gamma_{l}(x). 
\label{EQ:bounds_def}
\eea
\eml
Here, the symbol $\inf$ indicates the greatest lower
bound~\cite{FNOTE:inf} for a set of real numbers.  

It is convenient to write $\gamma_{l}(x)$ in terms of another function
$\beta_{l}(v)$, as follows:
\bml
\bea
\gamma_{l}(x) & = & \int_{0}^{\infty}\!\! d\theta \,\pi(\theta)
\,\beta_{l}(x/\sqrt{\theta}),
\label{EQ:bl_beta}
\\
\beta_{l}(v) & \equiv & \frac{v^{2}}{2 
\langle \theta^{-1} \rangle_{\pi} }
\nonumber \\
&&  -4 \int_{0}^{\infty} \!\!\! du \, \sqrt{u v} \, {\rm e}^{-(u^2
+ v^2)} I_{l-1}(2 u v).
\label{EQ:beta_def}
\eea
\eml
Here, we have used the definition of an average with respect to the
distribution $\pi$,
\beq 
\langle f(\theta) \rangle_{\pi} \equiv \int_{0}^{\infty}\!\! d\theta
\,\pi(\theta) \,f(\theta),
\label{EQ:moment_def}  
\eeq
and we need, in particular, the numerical value
\beq 
\langle \theta^{-1} \rangle_{\pi} \approx 0.881768, 
\label{EQ:moment_value}
\eeq
which can be obtained by using the function $\pi(\theta)$ of
Ref.~\cite{REF:epl}. 
As $\pi(\theta)$ is nonnegative and normalized to unity,
Eq.~(\ref{EQ:bl_beta}) implies that
\beq
\zeta \geq \bar{\gamma}_{|\sigma|} \geq \bar{\beta}_{|\sigma|},
\label{EQ:bound_beta}
\eeq
where
\beq
\bar{\beta}_{l} \equiv \inf_{v} \beta_{l}(v).
\label{EQ:bar_beta_def}
\eeq

The bounds for different values of the index $l$ are not independent. In fact,
because $I_{l}(x) < I_{l-1}(x)$ for $x > 0$ and $l \geq 1$, we have
the inequality $\eta^{(0)}_{l+1}(x,x') < \eta^{(0)}_{l}(x,x')$ for $x
x' > 0$ and $l \geq 1$, and from this inequality it follows that all
the bounds defined so far [$\gamma_{l}(x)$, $\bar{\gamma}_{l}$,
$\beta_{l}(v)$, and $\bar{\beta}_{l}$] are {\em increasing} functions
of $l$ for $l \geq 1$. Thus, if we obtain a positive lower bound for
one value of $l \geq 1$, the same bound applies for all larger values
of $l$.

In order to obtain more concrete results, we need an explicit
expression for $\beta_{l}(v)$. In App.~(\ref{APP:bl}) we obtain
the exact expression
\bml
\bea 
\beta_{l}(v) & = & \frac{v^{2}}{2 \langle \theta^{-1} \rangle_{\pi} }
-2 \frac{\Gamma(\frac{l}{2}+\frac{1}{4})}{\Gamma(l)} 
v^{l-1/2} M \left( \frac{l}{2}-\frac{1}{4},l,-v^2 \right),
\nonumber \\
\label{EQ:beta_comp} 
\eea
along with the asymptotic forms $\beta^{(>)}(v)$ and $\beta^{(<)}(v)$,
given by 
\bea
\beta_{l}(v) & \sim \beta^{(>)}(v) \equiv 
\frac{\displaystyle v^{2}}{\displaystyle 2 \langle \theta^{-1}
\rangle_{\pi} } & \, - \, 2  
\nonumber \\
&&
\mbox{ for $v \gg 1$},
\label{EQ:beta_v_big}
\\
\nonumber
\\
\beta_{l}(v) & \sim \beta^{(<)}(v) \equiv 
\frac{\displaystyle v^{2}}{\displaystyle 2 \langle \theta^{-1}
\rangle_{\pi} } &  
\, - \, 2 \, \frac{\Gamma(\frac{l}{2}+\frac{1}{4})}{\Gamma(l)} 
\, v^{l-1/2} 
\nonumber \\
&&
\mbox{ for $v \ll 1$},
\label{EQ:beta_v_small}
\eea
\eml
as well as the lower bounds
\bml
\bea
\beta_{l}(v) & > \beta^{(>)}(v) 
\qquad \qquad 
& \mbox{for $l > 1$},
\label{EQ:beta_bound_big}
\\
\beta_{l}(v) & \geq \beta^{(<)}(v) 
\qquad \qquad 
& \forall l.
\label{EQ:beta_bound_small}
\eea
\eml
Here $\Gamma(z)$ is the Gamma function, and $M(a,b,z)$ is a
confluent hypergeometric function~(\cite{REF:abramowitz}, Chap.~13)

As mentioned above, we need to show that $1 + \zeta$ is positive.
Thus, the quantity of interest is really $1 + \beta_{l}(v)$, as
opposed to $\beta_{l}(v)$. In Fig.~\ref{FIG:beta_v} we plot $1 +
\beta_{l}(v)$ as a function of $v$ for $0 \le l \le 4$, together with
its asymptotic form $1 + \beta^{(>)}(v)$ valid for large values of the
argument $v$.

%
\begin{figure}[htbp]
\centerline{\psfig{figure=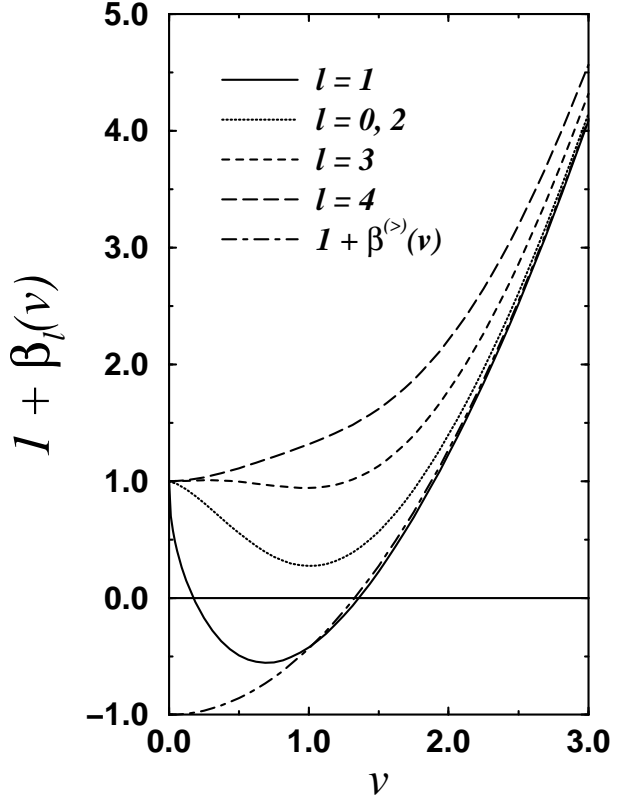,height=9.5cm}}
\vspace{1cm}
\nopagebreak
\narrowtext 
\centerline{\caption{Plot of $1 + \beta_{l}(v)$ (for $0 \le l \le 4$)
and $1 + \beta^{(>)}(v)$ as functions of $v$ \label{FIG:beta_v}}
}
\end{figure}
%

Let us now obtain the lower bounds $1 + \bar{\beta}_{l}$ for $1 +
\zeta$, and show that they are positive for $l \neq 1$.  For $l = 4$
(and, as $\beta_{l}(v)$ grows with $l$, for all $l \geq 4$),
$\beta_{l}(v)$ is positive for all nonzero $v$, and thus
$\bar{\beta}_{l}= \beta_{l}(0) = 0$. For $0 \leq l \leq 3$,
$\bar{\beta}_{l}$ is obtained by numerically minimizing
Eq.~(\ref{EQ:beta_comp}). In Table~\ref{TAB:lower_bounds}, we give the
numerical values for these bounds. These lower bounds establish that
all of the $\kappa_{lr}(\tilde{\bf k})$ are positive for $l \neq 1$.

\narrowtext
\begin{table}[htb]
\caption{Lower bounds for eigenvalues of the
Hessian.\label{TAB:lower_bounds}} 
\begin{tabular}{rcccl}
~~~ & $l$ & ~~~ & $1 + \bar{\beta}_{l}$ &~~~ \\
\tableline
~~~ & $1$ & ~~~ & $-0.55571$ &~~~ \\
~~~ & $0$, $2$ & ~~~ & $0.27376$ &~~~ \\
~~~ & $3$ & ~~~ & $0.94274$ &~~~ \\
~~~ & $\geq 4$ & ~~~ & $1$ &~~~ \\
\end{tabular}
\end{table}

Let us now focus on the remaining sector, namely $l = 1$. As, for this
case, our lower bound is negative, we can not draw any conclusion from
it. We have already shown that there is a zero mode, but there could
still be one or more negative eigenvalues, which would render the
proposed amorphous solid 
state unstable. The numerical solution of the radial
equation~(\ref{EQ:eigen_scaled_0}) for this case yields, within
numerical error~\cite{FNOTE:numerical_error}, the following two lowest
eigenvalues (both nondegenerate):
\bea
1 + \zeta_{10} & = & -0.00002 \pm  0.00005 
\nonumber \\
1 + \zeta_{11} & = & {\phantom -}0.98412 \pm  0.00009
\label{EQ:numerical_eigenvalues}
\eea
Evidently, $1 + \zeta_{10}$ corresponds to the expected zero mode, and we can
conclude that there are no further zero modes and that all other
eigenvalues are positive definite. 

To summarize, we have shown that $1 + \zeta \ge 0$ for any eigenvalue
$\zeta$ of Eq.~(\ref{EQ:eigen_scaled_0}), \ie, that all elements of
${\cal S}_{r}$ are positive or zero. In the next subsection we will
show that ${\cal S}_{r}$ and ${\cal S}_{o}$ are identical, and
therefore that all of the eigenvalues of the original Hessian are
either positive or zero.

\subsection{The one replica sector and the spurious eigenvalue}
\label{SEC:1rs}

We now need to return to the issues that we postponed earlier, namely
our extending of the Hessian matrix defined by Eq.~(\ref{EQ:F2_hrs})
so that it also be defined in the one replica sector, and the
differences between the spectra of the radial
Eqs.~(\ref{EQ:eigen_scaled_n}) and (\ref{EQ:eigen_scaled_0}).

As the Hessian matrix (both in its ``original'' and its ``extended''
versions) leaves exactly uncoupled fluctuations with different values
of $\tilde{\bf k}$, it is consistent to consider separately the MTI
fluctuations (those with $\tilde{\bf k} = {\bf 0}$) and the non-MTI
fluctuations (those with $\tilde{\bf k} \neq {\bf 0}$).

For the case of non-MTI fluctuations we will show that, in the limit
$n \to 0$, the hrs and 1rs are not coupled by the extended Hessian
matrix. Furthermore, each one of the eigenvectors belongs to one of
the sectors, inasmuch as it has a vanishing overlap (as $n \to 0$)
with vectors in the other sector.

To understand this issue, we need to look at the form that the 1rs
and hrs take, in the replica limit. For a wavevector 
\beq 
\hat{p} = ({\bf 0}, \ldots, {\bf 0}, {\bf p}, {\bf 0}, \ldots, {\bf
0})
\label{EQ:p_1rs}
\eeq
in the one replica sector, we have  
\beq
\tilde{\bf p} = {\bf p} \qquad \mbox{and} \qquad {\hat{p}}^{2} = {\bf
p}^{2}, 
\label{EQ:cond_1rs}
\eeq
and, by using Eq.~(\ref{EQ:scalar_product}) with $\hat{v} = \hat{w} =
\hat{p}$, we have
\beq
|{\breve{p}}| 
= \sqrt{ {\hat{p}}^{2} - \frac{{\tilde{\bf p}}^{2}}{1+n} } 
= \sqrt{ {\bf p}^{2} - \frac{{\bf p}^{2}}{1+n} } 
= \sqrt{\frac{n}{1+n}} |{\bf p}|.
\label{EQ:1rs_radial}
\eeq 
This means that the radial coordinate $|\breve{p}|$ goes to
zero like $n^{1/2}$ in the replica limit, $n \to 0$.
Moreover, for each fixed value of $\tilde{\bf p}$ ($ = {\bf p}$), the $n+1$
wavevectors defined by
\bml
\bea
\hat{e}_{\alpha}({\bf p}) & \equiv & ({\bf e}^{0},{\bf e}^{1},\cdots,{\bf
e}^{n}),
\\
{\bf e}^{\beta} & \equiv & 
\left\{ 
\begin{array}{ll}
{\bf 0} & \mbox{for $\beta \neq \alpha$}, \\
{\bf p} & \mbox{for $\beta = \alpha$},
\end{array}
\right.
\label{EQ:e_alpha}
\eea
\eml
with $\alpha = 0,\ldots,n$, are the only vectors in the one replica sector
that satisfy the condition that the sum of their $(n+1)$
component $d$-dimensional wavevectors is equal to ${\bf p}$. 
These two results tell us that the one replica sector corresponds, 
for fixed $\tilde{\bf p}$, 
to a set of $n + 1$ points that, in the
replica limit, converge to the origin of $\breve{p}$ space. 
Consequently, to see whether or not a given eigenvector has any
overlap with the 1rs, one needs to look at the properties of the
corresponding radial eigenfunction near the origin. 

Let us then consider the scaled radial equation
(\ref{EQ:eigen_scaled_n}) for the region close to the origin, and
let us keep $n > 0$ for the moment. By using the small-argument
behavior of the modified Bessel function (valid for $\nu \neq -1, -2,
\ldots$), 
\beq
I_{\nu}(z) \approx \frac{(z/2)^{\nu}}{\Gamma(\nu+1)},
\label{EQ:Bessel_z_small}
\eeq
we obtain the asymptotic form of the kernel $\eta$ for $x \ll 1$ and $x'
\alt 1$: 
\bml
\bea
\eta^{(n)}_{l}(x,x') & \approx & 2 
\,\frac{x^{l+(nd-1)/2}}{\Gamma(l+nd/2)} \,m_{l}(x'),
\label{EQ:eta_x_small}
\\
m_{l}(y) & \equiv &
\int_{0}^{\infty} \!\!
d\theta \, \pi(\theta) \,\theta^{-l}
\,{\rm e}^{-y^2/{\theta}} \,y^{l+(nd-1)/2}. 
\label{EQ:ml_def}
\eea
\eml
By inserting this asymptotic form into Eq.~(\ref{EQ:eigen_scaled_n}),
we obtain 
\bml
\bea
(\zeta - \frac{x^2}{2}) u(x) & \approx & \frac{-4 x^{l+(nd-1)/2}}
{\Gamma(l+nd/2)} \,U_{l},
\label{EQ:eqn_x_small}
\\
U_{l} & \equiv & \int_{0}^{\infty} \!\!dy \,\,m_{l}(y) \,u(y).
\label{EQ:Ul_def}
\eea
\eml
For $n$ positive and small, Eq.~(\ref{EQ:eqn_x_small}) can only be
satisfied for $\zeta \neq 0$~\cite{FNOTE:zeta_nonzero}.
The term proportional to $x^2$ is thus negligible, and we obtain, for
$x \ll 1$,
\beq
u(x) \approx \frac{-4 \,U_{l}}{\Gamma(l+nd/2)} 
\,\frac{x^{l+(nd-1)/2}}{\zeta}.  
\label{EQ:u_x_small}
\eeq
The leading behavior of this radial eigenfunction for $nd$ small and
positive depends on the value of the degree $l=|\sigma|$ of the
surface harmonic function.  For $l = 0$ there is one eigenfunction that
diverges at the origin like $x^{(nd-1)/2}$. Its eigenvalue $\zeta_{-}$
is given by the expression
\beq
\zeta_{-} \approx -\frac {4 \int_{0}^{\infty} \!\!dy \int_{0}^{\infty}
\!\! d\theta \, \pi(\theta) \,{\rm e}^{-y^2/{\theta}} \,y^{nd-1}} 
{\Gamma(nd/2)}, 
\label{EQ:zeta_x_small}
\eeq 
which, in the replica limit, becomes 
\beq
\lim_{n \to 0} \zeta_{-} = -2.
\label{EQ:zeta_lim}
\eeq 
The presence of this divergent eigenfunction as a solution of
Eq.~(\ref{EQ:eigen_scaled_n}) depends crucially on the singularity of
$\eta^{(n)}_{0}(x,x')$ at the origin for $n$ small but positive.  It
is {\em not} a solution of Eq.~(\ref{EQ:eigen_scaled_0}), which only
has solutions that go to zero at the origin: by Eq.~(\ref{EQ:0_2}), the radial equation for $n = 0$ and $l = 0$,
is identical to the radial equation for $n = 0$ and $l = 2$, which
means, by Eq.~(\ref{EQ:u_x_small}), that the radial
eigenfunction satisfies the condition 
\beq 
|u(x)| \alt x^{3/2} .
\label{EQ:u_reg}
\eeq

We will show below that the divergent eigenfunction present for $n >
0$ corresponds, in the replica limit, to an unphysical fluctuation,
\ie, a fluctuation in the one replica sector. In fact, from
Eqs.~(\ref{EQ:eig_gen}) and (\ref{EQ:zeta_lim}), we see that its
eigenvalue $\kappa_{-}(\tilde{\bf k})$ is negative for small
$\tilde{\bf k}$: \beq \kappa_{-}(\tilde{\bf k}) = -\epsilon +
\frac{\tilde{\bf k}^{2}}{2}.
\label{EQ:kappa_spurious}
\eeq
However for $l = 0$, all other radial eigenfunctions are orthogonal to
the one just found, and thus make the integral $U_{l}$ vanish. Their
behavior is controlled by the next power in the expansion of
$\eta^{(n)}_{0}(x,x')$, and consequently they vanish for $x \to 0$ at
least as fast as $x^{(nd+3)/2}$.  Moreover, for $l > 0$, by
Eq.~(\ref{EQ:u_x_small}) we see that all the radial eigenfunctions
vanish for $x \to 0$ as $x^{l+(nd-1)/2}$ or faster.

These results can be summarized as follows: for $n \to 0^{+}$ all but
one of the radial eigenfunctions are regular at the origin. The one
singular eigenfunction corresponds to $l=0$ and scales like
$x^{(nd-1)/2}$ for $x \ll 1$. The regular eigenfunctions can have any
value of $l$ and vanish for $x \to 0$ as $x^{(|l-1|+1/2)}$ or faster.

In all cases in which the eigenfunction is regular at the origin, it
is permissible to take the limit $n \to 0$ in
Eq.~(\ref{EQ:eigen_scaled_n}). This is because these eigenfunctions
vanish at the origin fast enough that the integral term in
Eq.~(\ref{EQ:eigen_scaled_n}) does not pick up any extra contribution
from the singularity of $\eta^{(n)}_{l}(x,x')$ [which, by
Eqs.~(\ref{EQ:eta_x_small}) and (\ref{EQ:ml_def}), is at most of order
$nd (x x')^{(nd-1)/2}$].
Thus, the spectrum of eigenvalues of Eq.~(\ref{EQ:eigen_scaled_0}) is
the same as the limit of the spectrum of Eq.~(\ref{EQ:eigen_scaled_n})
when $n \to 0$, except that the spurious eigenvalue $\zeta_{-}$ is
absent in the former and present in the latter.

We now show that the one replica sector fluctuations decouple from the
higher replica sector fluctuations in the replica limit.  Consider the
following complete orthonormal basis set for the fluctuations in the
one replica sector with fixed ${\bf \tilde{p}} = {\bf p}$:
\bea
w_{j}(\hat{k}) & \equiv & \sum_{\alpha=0}^{n} w_{j,\alpha}
\, \delta_{\hat{k}, \hat{e}_{\alpha}({\bf p})} ,
\nonumber \\
w_{j,\alpha} & \equiv & \frac{V^{n/2}}{\sqrt{n+1}} 
\, {\rm e}^{i 2\pi j \alpha/(n+1)}, 
\label{EQ:basis_1rs}
\eea
where $j=0,\ldots,n$. 

Let us compute the scalar product $\langle w_{j} | \psi_{r {\bf
\tilde{p}} \sigma} \rangle$ of one of these basis functions for the one
replica sector with one of the eigenfunctions of the extended Hessian,
which has the general form given in Eq.~(\ref{EQ:eigenfunction_prod}):
\bea
\lefteqn{ 
\langle w_{j} | \psi_{r {\bf \tilde{p}} \sigma} \rangle = 
\frac{1}{V^{n}} \sum_{\hat{k} \neq \hat{0}} w_{j}^{\ast}(\hat{k}) \, \psi_{r {\bf
\tilde{p}} \sigma}(\hat{k}) 
} 
\nonumber \\
&& \qquad 
= \frac{1}{V^{n}} \sum_{\alpha=0}^{n} w_{j,\alpha}^{\ast}
\Big[ 
(1\!+\!n)^{d/4} (2 \pi)^{nd/2} S_{\sigma}(\phi_{\breve{e}_{\alpha}(\bf{p})})
\nonumber \\
&& \; \qquad 
\times 
\big({\scriptstyle \sqrt{\frac{n}{1+n}} }|{\bf p}|\big)^{\frac{1-nd}{2}}
R_{r}\big({\scriptstyle \sqrt{\frac{n}{1+n}}} |{\bf p}|\big) 
\Big].
\label{EQ:1rs_psi}
\eea
Here, we have made use of the relation Eq.~(\ref{EQ:1rs_radial}). There
are two possible cases to consider, depending on whether or not $R_{r}$ is
singular at the origin. If it is singular, we have $l=0$ and, for
small $k$,
\beq
R(k) = \epsilon^{-1/4} u(k/\sqrt{\epsilon}) 
\approx {\cal N} \epsilon^{-nd/4} k^{(nd-1)/2},
\label{EQ:R_sing}
\eeq
where ${\cal N}$ is a normalization constant determined by
Eq.~(\ref{EQ:R_norm}). Its value is given by 
\beq
{\cal N} = \sqrt{nd} \big(1 + {\cal O}(nd) \big).
\label{EQ:N_n}
\eeq
As $l=0$, the angular part of $\psi_{r {\bf \tilde{p}} \sigma}$ is
isotropic, and is given by 
\beq
S_{0}(\phi) = \sqrt{\frac{1}{\tau_{nd}}} = \sqrt{\frac{\Gamma(nd/2)}
{2\pi^{nd/2}}} = (nd)^{-1/2} \big(1 + {\cal O}(nd) \big),
\label{EQ:S0_n}
\eeq
where $\tau_{nd} = {2\pi^{nd/2}}/{\Gamma(nd/2)}$ is the surface area
of a 
unit sphere in $nd$-dimensions. 
By combining Eqs.~(\ref{EQ:1rs_psi}), (\ref{EQ:R_sing}),
(\ref{EQ:N_n}), and (\ref{EQ:S0_n}), we obtain
\bea
\langle w_{j} | \psi_{r {\bf \tilde{p}} \sigma} \rangle 
& = & \sum_{\alpha=0}^{n} w_{j,\alpha}^{\ast} \big(1 + {\cal O}(n) \big) 
= \langle w_{j} | w_{0} \rangle \big(1 + {\cal O}(n)\big) 
\nonumber \\
& = & \delta_{j,0} \big(1 + {\cal O}(n) \big)
\label{EQ:couple}
\eea
This result implies that, in the limit $n \to 0$, the eigenfunction
that is singular at the origin lies entirely in the one replica sector.

Let us now consider the case in which $R_{r}$ is not singular at the
origin. In this case, for small $k$, the radial eigenfunction has the
form
\beq
R(k) = \epsilon^{-1/4} \, u(k/\sqrt{\epsilon}) 
\alt {\cal N} \epsilon^{-(|l-1|+1)/2} k^{(|l-1|+1/2)},
\label{EQ:R_reg}
\eeq
where the normalization constant ${\cal N}$ does not vanish in the
replica limit.
As, in this regular case, $l$ need not be zero, we have to obtain an
estimate for the normalization constant of the surface harmonic for
all values of $l$.  Consider a monomial $M_{m}(\phi)$ defined on the
$D$-dimensional unit sphere
\beq
M_{m}(\phi) \equiv \phi_{1}^{m_1} \cdots \phi_{D}^{m_D} = \frac{ x_{1}^{m_1}
\cdots x_{D}^{m_D} }{r^{m_1+ \cdots + m_D}}.
\label{EQ:monomial}
\eeq
Here $(x_{1},\ldots,x_{D})$ are the cartesian coordinates of a point $x$,
$r \equiv (x_{1}^{2} + \cdots + x_{D}^{2})^{1/2}$ is the radial coordinate
for the same point, and $ \phi \equiv (\phi_{1},\ldots,\phi_{D}) \equiv x/r$ is
the unit vector pointing in the direction of $x$. The integral of the
monomial over the unit sphere is
\bea 
\lefteqn{ \int d\phi \,M_{m}(\phi)  = \frac{\displaystyle \int d^{D}x
\,\,x_{1}^{m_1} 
\cdots x_{D}^{m_D} \,\,{\rm e}^{-(x_{1}^{2} + \cdots + x_{D}^{2})}}
{\displaystyle \int_0^{\infty} \!\!dr \,\,r^{D-1} 
\,\,r^{m_1+ \cdots + m_D}
\,\,{\rm e}^{-r^{2}} }
}
\nonumber \\
&& \qquad \qquad =
\left\{ 
\begin{array}{ll}
\frac{\displaystyle 2 \prod_{j=1}^{D} \Gamma \Big(\frac{1+m_{j}}{2} \Big)}
{\displaystyle \Gamma\left(\frac{D + \sum_{j=1}^{D} m_{j}}{2}\right) }
& \mbox{if $m_{j}$ even $\forall j$,} \\
~ & ~ \\
0 & \mbox{otherwise.}
\end{array}
\right.
\nonumber \\
&& 
\label{EQ:norm_int}
\eea
In the case of interest to us $D = nd$ and $\sum_{j=1}^{D} m_{j} =
2 |\sigma|$. {}From Eq.~(\ref{EQ:norm_int}) we conclude that the
normalization factor 
$N_{\sigma}$ for the surface harmonic $S_{\sigma}$ has, in the $n \to
0$ limit, the asymptotic form
\beq
N_{\sigma} \sim \sqrt{ \Gamma \left(\frac{nd}{2} + |\sigma| \right) } \sim 
\left\{ 
\begin{array}{ll}
n^{-1/2} & \mbox{for $|\sigma| = 0$,} \\
n^{0} & \mbox{for $|\sigma| \neq 0$.}
\end{array}
\right.
\label{EQ:N_l}
\eeq
Here, we have ignored factors that have finite limits when $n \to 0$.
This result can be summarized as follows
\bea
S_{\sigma}(\phi) & \sim & n^{\upsilon(\sigma)},
\nonumber \\
\upsilon(\sigma) & = & 
\left\{ 
\begin{array}{ll}
{-1/2} & \mbox{for $|\sigma| = 0$,} \\
{0} & \mbox{for $|\sigma| \neq 0$.}
\end{array}
\right.
\label{EQ:S_n}
\eea

By inserting Eqs.~(\ref{EQ:R_reg}) and (\ref{EQ:S_n}) into
Eq.~(\ref{EQ:1rs_psi}), we obtain the following scaling with $n$ for
the sought scalar product:
\bea
| \langle w_{j} | \psi_{r {\bf \tilde{p}} \sigma} \rangle |
& \alt &
(\sqrt{n}|\tilde{\bf k}|)^{1/2}
(\sqrt{n}|\tilde{\bf k}|)^{1/2 + |l - 1|}
\,n^{\upsilon(\sigma)},
\nonumber \\
& \alt &
\left\{ 
\begin{array}{ll}
n^{1/2} & \mbox{for $l = 0$,} \\
n^{(|l - 1| + 1)/2} & \mbox{for $l \neq 0$.}
\end{array}
\right.
\label{EQ:psi_n}
\eea
This relation shows that in the limit $n \to 0$ those radial
eigenfunctions that are regular at the origin give rise to
eigenvectors that lie entirely in the higher replica sector. 

For completeness, we now also compute explicitly the matrix elements of the
extended Hessian between members of the basis set
$\{w_{j}\}_{j=0}^{n}$ for the one replica sector fluctuations with
$\tilde{\bf k} = {\bf p}$,
\bea
\lefteqn{ \langle w_{m} | H | w_{j} \rangle  =  \frac{1}{V^{2n}}
\sum_{\hat{k},\hat{l} \neq \hat{0}} w_{m}^{\ast}(\hat{k}) \,
H_{\hat{k},\hat{l}} \,\, w_{j}(\hat{l}) } 
\nonumber \\
&& \quad = \frac{1}{V^{n} (1+n)} \Big[ \sum_{\alpha=0}^{n}
\Big( \epsilon + \frac{{\bf p}^{2}}{2} \Big) 
w_{m,\alpha}^{\ast} \, w_{j,\alpha}^{\phantom \ast} 
\nonumber \\
&& \quad
-2\epsilon \int_{0}^{\infty} \!\!d\theta \, \pi(\theta) 
\sum_{\alpha,\beta=0}^{n} \,{\rm e}^{{-[\hat{e}_{\alpha}({\bf
p})-\hat{e}_{\beta}({\bf p})]^{2}}/{\epsilon \theta}} 
w_{m,\alpha}^{\ast} \, w_{j,\beta}^{\phantom \ast} \Big]
\nonumber \\
&& \quad 
= \delta_{m,j} \Big( -\epsilon + \frac{{\bf p}^{2}}{2} \Big)
+ {\cal O}(n).
\label{EQ:eigen_1rs}
\eea  
Thus we see that, as expected, the eigenvalue obtained here is the same as
the one obtained in Eq.~(\ref{EQ:kappa_spurious}) for the singular
eigenfunction of the extended Hessian.

In summary, for non-MTI fluctuations, in the replica limit all regular
eigenfunctions of the extended Hessian are orthogonal to all of the
1rs vectors, and the singular eigenfunction of the extended Hessian
coincides with the isotropic ($j = 0$) fluctuation in the
1rs. Consequently, in the replica limit, the higher replica sector is
an invariant subspace for the extended Hessian, and therefore the
regular eigenfunctions of the extended Hessian are the eigenfunctions
of the original Hessian. More significantly, the eigenvalues of the
original Hessian are the eigenvalues of the extended Hessian for its
regular eigenfunctions.

For the case of MTI fluctuations, their components in the one replica
sector are exactly zero, because the conditions $\hat{k} \in$ 1rs and
$\tilde{\bf k} = {\bf 0}$ are incompatible. However, all of the
argument presented above still holds, except that now the radial
eigenfunction that is singular at the origin coincides with a spurious
fluctuation in the zero replica sector (\ie, a fluctuation of
$\Omega_{\hat{0}}$)~\cite{FNOTE:spurious_0rs}. The spectrum of the
original Hessian is, also in this case, given by the eigenvalues
corresponding to radial eigenfunctions regular at the origin.
Thus, we have shown that the sets ${\cal S}_{r}$ and ${\cal S}_{o}$
are identical, and we can use the results of Secs.~\ref{SEC:zero_mode}
and~\ref{SEC:lower_bounds} to characterize the spectrum of the
original Hessian.

The eigenvalues of the original Hessian have the general form
\beq
\kappa_{lr}(\tilde{\bf k}) = (1 + \zeta_{lr}) \epsilon 
+ \frac{{\tilde{\bf k}}^{2}}{2},
\label{EQ:eig_gen2}
\eeq
with 
\beq 
1 + \zeta_{10} = 0
\label{EQ:z10}
\eeq
and 
\beq 
1 + \zeta_{lr} > 0 \qquad \qquad \mbox{for $(l,r) \neq (1,0)$}.
\label{EQ:zlr}
\eeq
Therefore there is a zero mode corresponding to $(l,r) = (1,0)$ and
$\tilde{\bf k}={\bf 0}$, which is continued by a branch of soft
modes with eigenvalues
\beq
\kappa_{10}(\tilde{\bf k}) = \frac{{\tilde{\bf k}}^{2}}{2} \qquad
\mbox{($> 0$ for $\tilde{\bf k} \neq {\bf 0}$).}
\label{EQ:eig_10}
\eeq
All other eigenvalues are positive, with one continuous branch of
modes labeled by $\tilde{\bf k}$ for each value of $(l,r)$. The minimum
eigenvalue for each branch is given by 
\beq
\kappa_{lr}({\bf 0}) = (1 + \zeta_{lr}) \epsilon > 0,
\label{EQ:eig_min}
\eeq
which goes to zero as the transition is approached (\ie~as
$\epsilon \to 0$).
Consequently, the amorphous solid state of Ref.~\cite{REF:epl} is
locally stable near the transition.

\section{Randomly crosslinked macromolecules}
\label{SEC:rcms} 

We now consider one example of a semi-microscopic theory that exhibits
the amorphous solidification transition, namely the case of randomly
crosslinked linear macromolecules. 
In this theory there appears a control parameter $\mu ^{2}$ ($\equiv 1 +
\epsilon/3$) 
that determines the crosslink density, and such
that the system exhibits the liquid phase for $\mu ^{2} < 1 $ and the
amorphous solid phase for $\mu ^{2} > 1 $. In this semi-microscopic
theory, the field $\Omega_{\hat{p}}$ (with $\hat{p}$ in the one
replica sector) is present and allowed to fluctuate, and there is a coupling
parameter $\tilde{\lambda}_{n}^{2} \equiv \lambda^{2} -
\mu ^{2} \frac{V}{N} \frac{1}{V^{n}} $ associated with its
fluctuations. (The parameter $\lambda^{2}$ gives the strength of the
excluded-volume interaction between the macromolecules.) 
The free-energy functional (per macromolecule) has the
form\cite{REF:prl_1987,REF:cross} 
\begin{eqnarray}
\lefteqn{
nd{\cal F}_{n}(\{\Omega_{\hat{k}}\}) 
= \tilde{\lambda}_{n}^{2} \frac{N}{V}
\tilde{\sum}^{\dagger}_{\hat{p}} \vert \Omega_{\hat p} \vert^{2} 
+ \frac{\mu ^{2}}{V^{n}} {\overline{\sum}}^{\dagger}_{\hat{k}}  
\vert \Omega_{\hat k} \vert^{2}
} \nonumber \\  &&  \qquad 
- \ln{ \left\langle 
\exp \left(   
i \tilde{\lambda}_{n}^{2} \frac{2 N}{V}
\tilde{\sum}^{\dagger}_{\hat{p}} \real \Omega_{\hat p}
\rho^{\ast}_{\hat p}
\right. \right. }
\nonumber \\  &&  \qquad  \qquad  \qquad
{ \left. \left.
+ \frac{2 \mu ^{2}}{V^{n}} {\overline{\sum}}^{\dagger}_{\hat{k}}  
\real \Omega_{\hat k} \rho^{\ast}_{\hat k}
\right) 
\right\rangle^{W}_{n+1}},
\label{EQ:F_Omega}
\end{eqnarray}
where the symbol $\tilde{\sum}_{\hat{p}}$ denotes a sum over
replicated wave vectors in the one replica sector, and the $\dagger$
symbol additionally restricts any summation to the half space of
relevant wave vectors [i.e., d-dimensional or (n+1)d-dimensional] such
that their scalar product with a fixed unit vector (${\bf n}$ or
$\hat{n}$) is positive.  Here, we have used the definition of the
one-macromolecule Fourier transformed density $\rho_{\hat k} \,$, \ie,
\begin{equation}
\rho_{\hat k} \equiv \int_{0}^{1} \!\!ds\,\, \exp{i {\hat k} \cdot 
{\hat c}(s) },
\label{EQ:rho_def}
\end{equation}
for a macromolecular configuration ${\hat c}(s)$, and the 
Wiener replicated average $\left\langle \cdots
\right\rangle^{W}_{n+1}$ is defined by
\begin{equation}
\left\langle O \right\rangle^{W}_{n+1} \equiv \frac{
\int {\cal D}{\hat c} \,\, O 
\,\, \exp\Big\{-\frac{1}{2} \int_{0}^{1} \!\! ds
\left\vert \frac{d{\hat c}(s)}{{ds}} \right\vert^{2}
\Big\} 
}{
\int {\cal D}{\hat c} \,
\,\, \exp\Big\{-\frac{1}{2} \int_{0}^{1} \!\! ds
\left\vert \frac{d{\hat c}(s)}{{ds}}\right\vert^{2}
\Big\} 
}.
\label{EQ:Wiener_def_n1}
\end{equation}
Let us note here that to leading order in $\epsilon$ the amorphous
solid stationary point in this 
theory is the same as in the Landau theory discussed above, \ie, it
is also described by Eqs.~(\ref{EQ:ord_par_scale}), (\ref{EQ:scpieq}),
and (\ref{EQ:pi_norm}).

We now expand the free energy functional to quadratic order around a
stationary point, and obtain its second derivatives with respect to the
fields $\{\Omega_{\hat{q}}\}$. In this section we use the notations
$H$ and $\bar{H}$ to refer to the exact Hessian for the microscopic
theory and the extended Hessian for the Landau theory, respectively.
For ${\hat{k}}$ and ${\hat{k}'}$ both in the higher replica sector we
have 
\bea
\lefteqn{ \frac{ \delta^{2} [nd{\cal F}_{n}] }  
{\delta \Omega_{\hat{k}} \delta \Omega_{-\hat{k}'}} =
H^{\rm hh}_{\hat{k},\hat{k}'} } 
\nonumber \\
&& = \frac{\mu ^{2}}{V^{n}} \left(\delta_{{\hat{k}},{\hat{k}'}} 
- \frac{\mu ^{2}}{V^{n}} \langle \rho_{-\hat{k}} \rho_{\hat{k}'}
\rangle_{n+1,c}^{W,\bar{\Omega}} \right)
\nonumber \\
&& = \frac{\mu ^{2}} {3} \Big[ \delta_{{\hat{k}},{\hat{k}'}} 
\Big(\epsilon \!+\! \frac{{\hat{k}}^{2}}{2} \Big) 
- \delta_{\tilde{\bf k},\tilde{\bf k}'} 2 \epsilon \int_{0}^{\infty}
\!\! d\theta \, \pi(\theta) {\rm
e}^{-({\hat{k}}-{\hat{k}'})^{2}/{\epsilon \theta}}
\Big] 
\nonumber \\
&& \qquad \qquad \qquad + {\cal O}(\epsilon^{2}) 
\nonumber \\
&& = \frac{1}{3} \,\bar{H}^{\rm hh}_{\hat{k},\hat{k}'} + {\cal
O}(\epsilon^{2}). 
\label{EQ:F2_hrs_micro}
\eea
For ${\hat{k}}$ in the higher replica sector and ${\hat{p}}$ in the
one replica sector we have
\bea
\lefteqn{ \frac{ \delta^{2} [nd{\cal F}_{n}] }  
{\delta \Omega_{\hat{k}} \delta \Omega_{-\hat{p}}} =
H^{\rm h1}_{\hat{k},\hat{p}} }
\nonumber \\
&& \quad = -i \tilde{\lambda}^{2}_{n} \frac{N \mu ^{2}}{V^{1+n}} 
\langle \rho_{-\hat{k}} \rho_{\hat{p}} \rangle_{n+1,c}^{W,\bar{\Omega}} 
\nonumber \\
&& \quad =  -i \frac{\tilde{\lambda}^{2}_{n}}{V} \frac{N \mu ^{2}}{3} 
\delta_{\tilde{\bf k},\tilde{\bf p}} 2 \epsilon \int_{0}^{\infty}
\!\! d\theta \, \pi(\theta) {\rm
e}^{-({\hat{k}}-{\hat{p}})^{2}/{\epsilon \theta}} 
+ {\cal O}(\epsilon^{2}) 
\nonumber \\
&& \quad = \frac{i \tilde{\lambda}^{2}_{n} N}{3 V}
\,\bar{H}^{\rm h1}_{\hat{k},\hat{p}} + {\cal O}(\epsilon^{2}). 
\label{EQ:F2_hrs_1rs}
\eea
Finally, for both ${\hat{p}}$ and ${\hat{p}'}$ in the one replica
sector we have
\bea
\lefteqn{ \frac{ \delta^{2} [nd{\cal F}_{n}] }  
{\delta \Omega_{\hat{p}} \delta \Omega_{-\hat{p}'}} =
H^{11}_{\hat{p},\hat{p}'} }
\nonumber \\
&& \qquad = \frac{\tilde{\lambda}^{2}_{n} N} {V} 
\Big( \delta_{{\hat{p}},{\hat{p}'}} 
+ \frac{\tilde{\lambda}^{2}_{n} N} {V} 
\langle \rho_{-\hat{p}} \rho_{\hat{p}'}
\rangle_{n+1,c}^{W,\bar{\Omega}} \Big) 
\nonumber \\
&& \qquad = \delta_{{\hat{p}},{\hat{p}'}}
\frac{\tilde{\lambda}^{2}_{n} N} {V} \Big\{ 1 
+ \frac{\tilde{\lambda}^{2}_{n} N} {V} 
[ 1 + {\cal O}(\epsilon) + {\cal O}({\hat{p}}^{2}) ] \Big\}. 
\label{EQ:F2_1rs}
\eea
In obtaining these formulas we have made use of the definition
\begin{equation}
\langle O \rangle^{W,\bar{\Omega}}_{n+1} \equiv 
\frac{
\left\langle   O
\exp \left(   
i \tilde{\lambda}_{n}^{2} \frac{N}{V}
\tilde{\sum}_{\hat{p}} \bar{\Omega}_{\hat p} \rho^{\ast}_{\hat p}
+ \frac{\mu ^{2}}{V^{n}} {\overline{\sum}}_{\hat{k}}  
\bar{\Omega}_{\hat k} \rho^{\ast}_{\hat k}
\right)
\right\rangle^{W}_{n+1}
} {
\left\langle   
\exp \left(   
i \tilde{\lambda}_{n}^{2} \frac{N}{V}
\tilde{\sum}_{\hat{p}} \bar{\Omega}_{\hat p} \rho^{\ast}_{\hat p}
+ \frac{\mu ^{2}}{V^{n}} {\overline{\sum}}_{\hat{k}}  
\bar{\Omega}_{\hat k} \rho^{\ast}_{\hat k}
\right) 
\right\rangle^{W}_{n+1}
}.
\label{EQ:WO_av_def} 
\end{equation}
The notations $H^{\rm hh}$, $H^{\rm h1}$, and  $H^{11}$, respectively,
refer to the higher replica, cross-sector, and one replica parts of
the Hessian matrix. Fig.~\ref{FIG:micro_H} depicts the relation
between $H$ and $\bar{H}$.

%
\begin{figure}[htbp]
\centerline{\psfig{figure=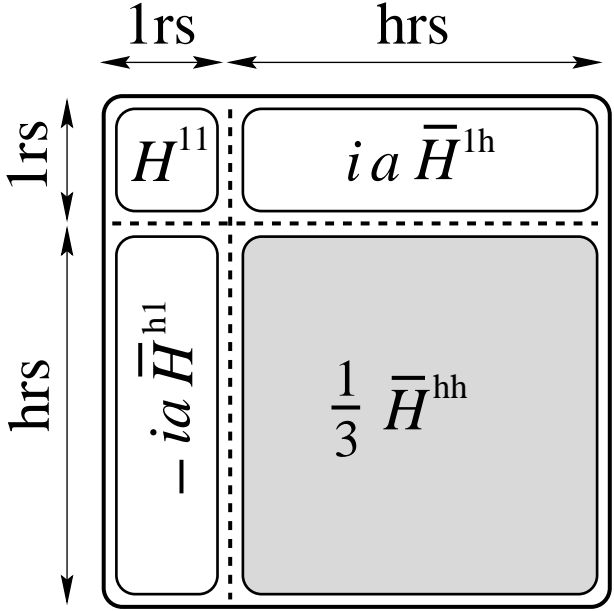,height=8.0cm}}
\vspace{1cm}
\nopagebreak
\narrowtext 
\centerline{\caption{
The Hessian matrix $H$ for the semimicroscopic theory in terms of the
extended Hessian matrix $\bar{H}$ for the Landau theory. Each block of $H$ is
written in terms of the corresponding block of $\bar{H}$, except for
$H^{\rm 11}$. ($H^{\rm 11}$ is nonzero when $\epsilon \to 0$, as
opposed to $\bar{H}^{\rm 11}$, which vanishes linearly with $\epsilon$
near the transition.) \label{FIG:micro_H}}
}
\end{figure}
%

As MTI (\ie~$\tilde{\bf k} = {\bf 0}$) fluctuations do not have any
component in the 1rs, the 
relevant Hessian in this case is just $H^{\rm hh} = (1/3)\bar{H}^{\rm hh}$.
Consequently the results obtained for the Landau theory tell us that
there is a zero eigenvalue corresponding to the anticipated zero mode,
and that the remaining eigenvalues are positive.

Let us now consider general fluctuations.  We will show that, in the
replica limit, the eigenvectors of the Hessian $H$ for this problem
are the same as the eigenvectors of the extended Hessian $\bar{H}$ for
the Landau theory, and the one replica and higher replica sectors are
again invariant subspaces for this Hessian.

Let us consider a regular eigenvector $| \psi_{r {\bf
\tilde{p}} \sigma} \rangle$ of $\bar{H}$ and one of the elements of
the basis set $\{ | w_{j} \rangle \}_{j=0}^{n}$ of the one replica
sector fluctuations. By Eq.~(\ref{EQ:F2_1rs}), 
\bea
H^{11} | w_{j} \rangle & = & \kappa_{\mbox{1rs}} | w_{j} \rangle,
\nonumber \\
\kappa_{\mbox{1rs}} & = & \frac{\tilde{\lambda}^{2}_{n} N} {V} \Big\{ 1 
+ \frac{\tilde{\lambda}^{2}_{n} N} {V} 
[ 1 + {\cal O}(\epsilon) + {\cal O}({\hat{p}}^{2}) ] \Big\}, 
\label{EQ:kappa_1rs}
\eea
and therefore 
\beq
| \langle w_{j} | H^{11} | \psi_{r {\bf \tilde{p}} \sigma} \rangle | 
= | \kappa_{\mbox{1rs}} 
\langle w_{j} | \psi_{r {\bf \tilde{p}} \sigma} \rangle |
\alt {\cal O}(\sqrt{n}).
\label{EQ:H11_zero}
\eeq
Analogously, by Eq.~(\ref{EQ:eigen_1rs}), 
\beq
\bar{H}^{11} | w_{j} \rangle = \Big( -\epsilon + \frac{{\bf
p}^{2}}{2} \Big) | w_{j} \rangle + {\cal O}(n),
\label{EQ:barH_1rs}
\eeq
and 
\beq
| \langle w_{j} | \bar{H}^{11} | \psi_{r {\bf \tilde{p}} \sigma} \rangle | 
= \Big| \Big( -\epsilon + \frac{{\bf p}^{2}}{2} \Big) 
\langle w_{j} | \psi_{r {\bf \tilde{p}} \sigma} \rangle \Big|
\alt {\cal O}(\sqrt{n}).
\label{EQ:barH11_zero}
\eeq
By combining Eqs.~(\ref{EQ:F2_hrs_1rs}), (\ref{EQ:H11_zero}), and
(\ref{EQ:barH11_zero}), we can now estimate the matrix element:
\bea
\lefteqn{ 
| \langle w_{j} | H | \psi_{r {\bf \tilde{p}} \sigma} \rangle | 
= | \langle w_{j} | H^{11} + H^{1 \rm h} | \psi_{r {\bf \tilde{p}} \sigma}
\rangle |  
}
\nonumber \\
&& \qquad 
= | \langle w_{j} | H^{1 \rm h} | \psi_{r {\bf \tilde{p}} \sigma} \rangle
+ {\cal O}(\sqrt{n}) | 
\nonumber \\
&& \qquad 
= | \langle w_{j} | \frac{i \tilde{\lambda}^{2}_{n} N}{3 V}
\,\bar{H}^{1 \rm h} | \psi_{r {\bf \tilde{p}} \sigma} \rangle
+ {\cal O}(\sqrt{n}) | 
\nonumber \\
&& \qquad 
= \Big| \frac{i \tilde{\lambda}^{2}_{n} N}{3 V} \, 
\langle w_{j} | \bar{H}^{1 \rm h} + \bar{H}^{11} | \psi_{r {\bf
\tilde{p}} \sigma} \rangle 
+ {\cal O}(\sqrt{n}) \Big| 
\nonumber \\
&& \qquad 
= \Big| \frac{i \tilde{\lambda}^{2}_{n} N}{3 V} \, 
\langle w_{j} | \bar{H} | \psi_{r {\bf
\tilde{p}} \sigma} \rangle 
+ {\cal O}(\sqrt{n}) \Big| 
\nonumber \\
&& \qquad 
\alt {\cal O}(\sqrt{n}).
\label{EQ:1h_zero}
\eea
This means that, in the replica limit, $H | \psi_{r {\bf \tilde{p}}
\sigma} \rangle$ has no projection in the one replica sector, and also
that $H | w_{j} \rangle$ has no projection in the higher replica
sector. Therefore, also in this problem the one replica sector and the
higher replica sector are decoupled invariant subspaces of the Hessian
in the $n \to 0$ limit. In the one replica sector, the eigenvalue is
$\kappa_{\mbox{1rs}} > 0$. In the higher replica sector, as $H^{\rm
hh} = (1/3)\bar{H}^{\rm hh}$, the eigenvectors are the same as for the
Landau theory, and the eigenvalues are obtained from those in the
Landau theory by multiplying by $1/3$. As discussed before, all of
these eigenvalues are positive, except for a unique zero mode.  Thus,
also for the semi-microscopic theory of randomly crosslinked
macromolecules, the amorphous solid state of Ref.~\cite{REF:epl} is
locally stable near the transition.

\section{Summary and concluding remarks}
\label{SEC:conclusions}
In this Paper we have shown that in a system with random constraints
near the liquid--amorphous-solid transition, the amorphous solid state
of Ref.~\cite{REF:epl} is a locally stable thermodynamic
state~\cite{FNOTE:support}. In order to do this, we have examined the
eigenvalue spectra of the stability matrices, in the contexts of both
the Landau theory for the transition and a semi-microscopic model of
randomly crosslinked macromolecular systems. In both cases the
spectrum turns out to be non-negative, with only a single zero
eigenvalue, and all the others positive.

Let us remark that even though we {\em do\/} find a zero eigenvalue for
the stability matrix, we still declare that the stationary point is
locally {\em stable\/}, as opposed to locally {\em marginally
stable\/}. This is because in this system translational 
invariance is spontaneously broken, and therefore there is a manifold
of equivalent states that have exactly the same free energy and are
connected to each other by the continuous symmetries of the
system. The zero eigenvalue (a.k.a. Goldstone mode) simply indicates
that the free energy does not change if one applies an infinitesimal
translation to the thermodynamic state.

In close analogy to the phonon spectra of ordinary solids, the
fluctuation eigenvalues can be classified into two types: a soft
branch of modes associated with ``almost rigid'' displacements of the
whole system (analogous to the acoustic phonon branch), with eigenvalues
$\kappa_{10}(\tilde{\bf k}) = {{\tilde{\bf k}}^{2}}/2$, and a set of
stiff modes in which the structure of the system is altered more
strongly (analogous to the set of optical phonon branches), with
eigenvalues $\kappa_{lr}(\tilde{\bf k}) = \epsilon(1 + \zeta_{lr}) +
{{\tilde{\bf k}}^{2}}/2$. In addition, there is in our case a
softening of the system, because the eigenvalues of the stiff
modes go to zero at the transition.

We have only addressed the issue of the {\em local} stability of the
amorphous solid state. It is much harder to determine whether the
amorphous solid state is {\em globally} stable, as the order parameter
space to be explored is enormous. In particular, one could consider
the possibility of a replica-symmetry breaking saddle point also being
present and dominating the physical behavior of the
system~\cite{FNOTE:no_rsb}. However, there are strong indications
(although by no means conclusive evidence) that the (replica
symmetric) saddle point considered here is indeed {\em globally}
stable. These indications mainly come from molecular dynamics
simulations~\cite{REF:Plischke} as the solid state observed in the
simulations appears to be identical to the one proposed in
Ref.~\cite{REF:epl}.

An intriguing problem, left open for further study, is to
establish how the structure of the eigenvalue spectrum of the Hessian
matrix, and in particular the softening of the system near the
transition, manifest themselves in the dynamics of the system.

\section{Acknowledgments}
\label{SEC:acknowledgments}
We gratefully acknowledge support from the University of Illinois at
Urbana-Champaign (H.E.C.), from the U.S. NSF through Grant
No.~DMR99-75187 (P.M.G.), from NATO through CRG 94090 (H.E.C., P.M.G.,
A.Z.), and from the DFG through SFB 345 (A.Z.).

\appendix

\section{Non-diagonal matrix elements for the Hessian}
\label{APP:matrix_el}

In this Appendix we collect some useful information concerning surface
harmonic functions, and use it to compute the matrix elements of the
non-diagonal part $H^{O}$ of the Hessian in the basis
$\{\varphi_{p {\bf \tilde{p}} \sigma}\}$.

The Gegenbauer (also called hyperspherical) polynomials play a
role in $D$ dimensions and with regard to the surface
harmonics $S_{\sigma}$ analogous to the role Legendre polynomials play in $3$
dimensions and with regard to the spherical harmonics $Y_{lm}$. The 
Gegenbauer polynomial $C^{\nu}_{l}(x)$ of degree $l$ is defined by the
generating function (see, e.g., Ref.~\cite{REF:Bateman}, Vol.~II,
Sec.~11.1.2) 
\beq 
(1 -2xt + t^{2})^{-\nu} = \sum_{l=0}^{\infty} C^{\nu}_{l}(x) \, t^{l}.
\label{EQ:Cn_gen}
\eeq

There is a generalization to dimension $D \equiv p+2$ of the addition theorem
for spherical harmonics, which relates the Gegenbauer polynomial to a
sum of surface harmonics (see, e.g., Ref.~\cite{REF:Bateman}, Sec.~11.4):
\bea
C^{p/2}_{l}(\phi' \!\cdot \! \phi) 
& = & \frac{C^{p/2}_{l}(1) \, \tau_{D}}{h(l,p)} 
\sum_{|\sigma| = l} S^{\ast}_{\sigma}(\phi') \, S_{\sigma}(\phi) 
\nonumber \\
& = & \frac{4 \pi^{1+p/2} }{(2l+p) \Gamma(p/2)}
\sum_{|\sigma| = l} S^{\ast}_{\sigma}(\phi') \, S_{\sigma}(\phi).
\label{EQ:add_thm_D}
\eea
Here, $\phi'$ and $\phi$ are any unit $D$-dimensional vectors,
$|\sigma|$ is the degree of the surface harmonic $S_{\sigma}$ as a
trigonometric polynomial, $h(l,p)$
is the number of linearly independent surface harmonics of degree $l$
in dimension $p+2$,
and $\tau_{D} = 2 \pi^{D/2} / \Gamma(D/2)$ is the surface area of a 
$D$-dimensional unit sphere.  [As $C^{1/2}_{l}(x)$ is equal to the
Legendre polynomial $P_{l}(x)$, formula~(\ref{EQ:add_thm_D}) reduces,
for $D=3$, to the usual addition theorem.]

We also make use of the identity (see, e.g., Ref.~\cite{REF:Bateman},
Vol.~II, Sec.~7.15)
\beq
z^{\nu} {\rm e}^{x z} = 2^{\nu} \Gamma(\nu) \sum_{n=0}^{\infty}
(n+\nu) \, C^{\nu}_{n}(x) \, I_{n+\nu}(z),
\label{EQ:exp_Cn}
\eeq
where $I_{\nu}(z)$ is the modified Bessel function of order $\nu$.

In the case of dimension $D=nd$, by combining
Eqs.~(\ref{EQ:add_thm_D}) and (\ref{EQ:exp_Cn}), the following
identity is obtained 
\bea
\lefteqn{
\exp(x \phi' \!\cdot \! \phi) = 2 \pi^{nd/2} (x/2)^{1-nd/2}
}
\nonumber \\ 
&& \quad \times 
\sum_{l=0}^{\infty} I_{l-1+nd/2}(x) \sum_{|\sigma|=l}
S^{\ast}_{\sigma}(\phi') \, S_{\sigma}(\phi).
\label{EQ:exp_decomp}
\eea
Here, $x$ is any real number,  and $\phi'$ and $\phi$ are unit
$nd$-dimensional vectors.

Let us now compute the matrix elements of the non-diagonal
part $H^{O}$ of the Hessian in the basis $\{\varphi_{p {\bf
\tilde{p}} \sigma}\}$. By using Eqs.~(\ref{EQ:basis}) and
(\ref{EQ:F2_T}) we obtain
\bea
\lefteqn{ \langle \varphi_{p' {\bf \tilde{p}'} \sigma'} | H^{O}
| \varphi_{p {\bf \tilde{p}} \sigma} \rangle } \nonumber \\
&& = V^{2} \int 
\frac{d{\bf \tilde{k}} \, d{\bf \tilde{l}} \, d{\breve{k}} \, d{\breve{l}}}
{(1+n)^{d} (2 \pi)^{(1+n)2d}}
(1+n)^{d/2} (2 \pi)^{nd} 
{p'}^{\frac{1-nd}{2}} 
\nonumber \\
&&
\times 
\delta_{{\bf \tilde{p}'}{\bf \tilde{k}}} \,
\delta(|\breve{k}|-p') \, S^{\ast}_{\sigma'}(\phi_{\breve{k}})
\,\, 
p^{\frac{1-nd}{2}}
\delta_{{\bf \tilde{p}}, {\bf \tilde{l}}} \,
\delta(|\breve{l}|-p) \, S_{\sigma}(\phi_{\breve{l}})
\nonumber \\
&&
\times \, 
\delta_{{\bf \tilde{k}}, {\bf \tilde{l}}}
\,\, (- 2 \epsilon) 
\int_{0}^{\infty} \!\! d\theta \, \pi(\theta) {\rm
e}^{-({\breve{k}}-{\breve{l}}\,)^{2}/{\epsilon \theta}},
\label{EQ:HO_1} \\
&&
= 
\frac{- 2 \epsilon \,\, (2 \pi)^{-nd} }{ (1+n)^{d/2} }
\sum_{{\bf \tilde{k}},{\bf \tilde{l}}}
\delta_{{\bf \tilde{p}'}, {\bf \tilde{k}}} \,
\delta_{{\bf \tilde{p}}, {\bf \tilde{l}}} \,
\delta_{{\bf \tilde{k}}, {\bf \tilde{l}}} \,
\,\,
\nonumber \\
&&
\times \, 
\int_{0}^{\infty} \!\! \delta(|\breve{k}|-p') 
\, |\breve{k}|^{nd-1} d|\breve{k}| 
\int_{0}^{\infty} \!\! \delta(|\breve{l}|-p) 
\, |\breve{l}|^{nd-1} d|\breve{l}| 
\nonumber \\
&&
\times \, 
\,\, (p p')^{\frac{1-nd}{2}}
\int_{0}^{\infty} \!\! d\theta \, \pi(\theta) \, {\rm
e}^{-(p^{2}+{p'}^{2})/{\epsilon \theta}}
\nonumber \\
&&
\times 
\! \int \! d\phi_{\breve{k}} \int \! d\phi_{\breve{l}} 
\,\, S^{\ast}_{\sigma'}(\phi_{\breve{k}})
\, \exp(2 p p' \phi_{\breve{k}} \! \cdot  \! \phi_{\breve{l}} /
{\epsilon \theta}) 
\, S_{\sigma}(\phi_{\breve{l}}).
\label{EQ:HO_2} 
\eea
In the second step, we have separated the $\breve{k}$ and $\breve{l}$
integrals into radial and angular parts. The angular integrals can be
performed with the help of the identity~(\ref{EQ:exp_decomp}) and
by using the orthonormality of the surface harmonics to obtain 
\bea
\lefteqn{
\langle \varphi_{p' {\bf \tilde{p}'} \sigma'} | H^{O}
| \varphi_{p {\bf \tilde{p}} \sigma} \rangle = 
\delta_{{\bf \tilde{p}'}, {\bf \tilde{p}}} 
\, \delta_{\sigma', \sigma} 
\frac{ (-2 \epsilon) \epsilon^{(nd-1)/2} } { 2^{nd} \pi^{nd/2}
(1+n)^{d/2} }
}
\nonumber \\ 
&& \times 
2 \sqrt{{p p'}/\epsilon}
\int_{0}^{\infty} \!\! \frac{d\theta \, \pi(\theta)}{\theta^{1-nd/2}} \, 
{\rm e}^{-(p'^2 + p^2)/{\epsilon \theta}}  
\, I_{|\sigma|-1+nd/2} \Big(\frac{2 p' p}{\epsilon \theta}\Big),
\nonumber \\
\label{EQ:HO_res_app}
\eea
which is equivalent to Eqs.~(\ref{EQ:HO_res}), (\ref{EQ:cn_def}) and
(\ref{EQ:eta_def}).

\section{Radial equation for the zero mode}
\label{APP:radial_zero}
In this Appendix we show that the scaled radial function of
Eq.~(\ref{EQ:radial_fluct_scaled}) corresponding to a change in the
system due to a rigid displacement is a solution of the scaled radial
eigenfunction equation Eq.~(\ref{EQ:eigen_scaled_0}) with $\zeta =
-1$. 

Let us first consider the diagonal term. By inserting the explicit
form for $u(x)$, and then performing an integration by parts we obtain
\bea
\frac{x^2}{2} u(x) & = & \frac{\sqrt{x}}{2} \int_{0}^{\infty} \!\!
d\theta\,\pi(\theta) \, \theta^{2} \frac{d}{d\theta}({\rm
e}^{-x^{2}/\theta})\,
\nonumber \\
& = & - \frac{\sqrt{x}}{2} \int_{0}^{\infty} \!\!
d\theta\, {\rm e}^{-x^{2}/\theta} \frac{d}{d\theta}\{\theta^{2}
\pi(\theta)\} 
\nonumber \\
& = & - \sqrt{x}\int_{0}^{\infty} \!\!
d\theta\, {\rm e}^{-x^{2}/\theta} \{\frac{\theta^{2}}{2} \frac{d}{d\theta}
\pi(\theta) + \theta \pi(\theta)\}. 
\label{EQ:0_diag}
\eea

Now the non-diagonal term gives 
\bea
\lefteqn{-2 \int_{0}^{\infty} \!\!dx' \eta^{(0)}_{1}(x,x') \,\, u(x')}
\nonumber \\
& = & -2 \int_{0}^{\infty} \!\!dx' \, d{\theta} \, d{\theta'} 2 \sqrt{x x'} 
\frac{\pi(\theta)}{\theta} {\rm e}^{-(x^{2}+x'^{2})/\theta}
I_{0} \Big(\frac{2 x x'}{\theta}\Big) 
\nonumber \\
&& \qquad \times \sqrt{x'} \, \pi(\theta') \, {\rm e}^{-x'^{2}/\theta'}.
\label{EQ:0_ndiag_1}  
\eea
By making use of the identity~\cite{REF:Gradshteyn}
\beq
2 \int_{0}^{\infty} \!\!dx \, x \, {\rm e}^{-a x^{2}} I_{0}(b x) = \frac{
{\rm e}^{b^2/{4a}} }{a},
\label{EQ:I0_int}
\eeq
we perform the integration over $x'$ in Eq.~(\ref{EQ:0_ndiag_1}), thus
obtaining
\bea
\lefteqn{-2 \int_{0}^{\infty} \!\!dx' \eta^{(0)}_{1}(x,x') \,\, u(x')}
\nonumber \\
& = & -2 \sqrt{x} \int_{0}^{\infty} \!\!d{\theta}d{\theta'}   
\frac{\theta' \pi(\theta) \pi(\theta')}{\theta+\theta'} {\rm
e}^{-x^{2}/(\theta+\theta')}
\nonumber \\
& = & - \sqrt{x} \int_{0}^{\infty} \!\!
d\theta\, {\rm e}^{-x^{2}/\theta} 
\int_{0}^{\theta} d{\theta'} \pi(\theta) \pi(\theta').
\label{EQ:0_ndiag_2}
\eea

Finally, we combine Eqs.~(\ref{EQ:0_diag}) and (\ref{EQ:0_ndiag_2}) to
obtain
\bea
\lefteqn{ u(x) + \frac{x^{2}}{2} u(x) -2 \int_{0}^{\infty}
dx' \eta^{(0)}_{|\sigma|}(x,x') \,\, u(x')} \nonumber \\
& = & 
\sqrt{x} \int_{0}^{\infty} \!\!
d\theta\, {\rm e}^{-x^{2}/\theta} 
\Big\{ -\frac{\theta^{2}}{2} \frac{d}{d\theta} \pi(\theta) 
\nonumber \\
&&
+ (1 - \theta) \pi(\theta)
-\int_{0}^{\theta} d{\theta'} \pi(\theta) \pi(\theta') \Big\}
\nonumber \\
& = & 0.
\label{EQ:0_shown}
\eea
The justification of the last equality comes from the factor in braces being zero by the stationarity condition,
Eq.~(\ref{EQ:scpieq}). 

Thus we have shown that Eq.~(\ref{EQ:eigen_scaled_0}) is satisfied by
$u(x)$ with the eigenvalue $\zeta = -1$.

\section{Computation of lower bounds}
\label{APP:bl}
In this Appendix we study in detail the bound-function $\beta_{l}(v)$.
We decompose $\beta_{l}(v)$ as follows:
\bea
\beta_{l}(v) & = & \frac{v^{2}}{2 \langle \theta^{-1} \rangle_{\pi} } 
-2 j_{l}(v)  
\nonumber \\
j_{l}(v) & \equiv & 2 \int_{0}^{\infty} \!\!\! du \, \sqrt{u v} \, {\rm
e}^{-(u^2 + v^2)} I_{l-1}(2 u v)
\label{EQ:jl_def}
\eea
We now compute analytically the integral defining $j_{l}(v)$:
\bea
j_{l}(v) & = & \sqrt{v} \, {\rm e}^{-v^2} \int_{0}^{\infty} \!\!\! dy \,
y^{-1/4} {\rm e}^{-y} I_{l-1}(2 v \sqrt{y})
\nonumber \\
& = & 
\frac{\Gamma(l/2+1/4)}{\Gamma(l)} 
v^{l-1/2} e^{-v^2} M \left( \frac{l}{2}+\frac{1}{4},l,v^2 \right)
\nonumber \\
& = & 
\frac{\Gamma(l/2+1/4)}{\Gamma(l)} 
v^{l-1/2} M \left( \frac{l}{2}-\frac{1}{4},l,-v^2 \right),
\label{EQ:j_comp}
\eea
where $M(a,b,z)$ is a confluent hypergeometric
function~(\cite{REF:abramowitz}, Chap.~13). By inserting this
expression into Eq.~(\ref{EQ:jl_def}), we obtain
Eq.~(\ref{EQ:beta_comp}). 

We can obtain more information by using the following integral formula
for the confluent hypergeometric function~(\cite{REF:abramowitz},
Chap.~13), valid for $\real a > 0$ and $\real b > 0$:
\beq
\frac{\Gamma(b-a) \Gamma(a)}{\Gamma(b)} M(a,b,z) = 
\int_{0}^{1} {\rm e}^{z t} t^{a-1} (1-t)^{b-a-1} dt.
\label{EQ:M_integral}
\eeq
This implies that 
\beq
j_{l}(v) =
\frac{(v^2)^{\frac{l}{2}\!-\!\frac{1}{4}}}
{\Gamma(\frac{l}{2}\!-\!\frac{1}{4})}
\int_{0}^{1} {\rm e}^{-t v^{2}} t^{(\frac{l}{2}\!-\!\frac{1}{4})-1} 
(1-t)^{(\frac{l}{2}\!+\!\frac{1}{4})-1} dt. 
\label{EQ:j_integral}
\eeq
By considering  the  fact that the exponential in the integrand is
always less than or 
equal to $1$, this formula can immediately be bounded above, as follows:
\bea
j_{l}(v) & \leq &
\frac{(v^2)^{\frac{l}{2}\!-\!\frac{1}{4}}}{\Gamma(\frac{l}{2}\!-\!\frac{1}{4})}
\int_{0}^{1} t^{(l/2-1/4)-1} (1-t)^{(l/2+1/4)-1} dt 
\nonumber \\ & = & 
\frac{(v^2)^{\frac{l}{2}\!-\!\frac{1}{4}}}{\Gamma(\frac{l}{2}\!-\!\frac{1}{4})}
B \left( \frac{l}{2}\!-\!\frac{1}{4},\frac{l}{2}\!+\!\frac{1}{4} \right)  
= \frac{\Gamma(\frac{l}{2}\!+\!\frac{1}{4})}{\Gamma(l)} v^{l-1/2},
\nonumber \\
&& 
\label{EQ:j_leq1}
\eea
where $B(x,y) = \Gamma(x) \Gamma(y) / \Gamma(x+y)$ is the Beta
function~(\cite{REF:abramowitz}, Sec.~6.2).
By combining this inequality with Eq.~(\ref{EQ:jl_def}) we obtain the
bound stated in Eq.~(\ref{EQ:beta_bound_small}). Moreover, in the
limit $v \ll 1$, it is legitimate to replace the exponential factor in
Eq.~(\ref{EQ:j_integral}) by $1$ inside
the integral, and thus the same expression gives the asymptotic form
in the $v \ll 1$ regime, as quoted in Eq.~(\ref{EQ:beta_v_small}).

An additional bound can be obtained for $l>1$ by taking into account
the fact that the factor $(1-t)^{(l/2+1/4)-1}$ in the integrand is
less than or equal to unity, so that
\beq
j_{l}(v) <  
\frac{(v^2)^{\frac{l}{2}\!-\!\frac{1}{4}}}
{\Gamma(\frac{l}{2}\!-\!\frac{1}{4})}
\int_{0}^{\infty} {\rm e}^{-t v^{2}} t^{(l/2-1/4)-1} dt = 1.
\label{EQ:j_leq2}
\eeq
This gives the lower bound of Eq.~(\ref{EQ:beta_bound_big}).  When $v
\gg 1$, the same expression provides the asymptotic form for all values
of $l$, Eq.~(\ref{EQ:beta_v_big}), as, in that limit, the integral is
dominated by the region near the origin, where the factor
$(1-t)^{(l/2+1/4)-1}$ is close to unity.

\end{multicols}

\begin{references}
\bibitem{FNOTE:irrel_const}
A constant is ignored here that has to do with the symmetry under
permutations of the macromolecules. See Ref.~\cite{REF:cross},
Secs.~2.4 and 2.6 for a full discussion.
\bibitem{FNOTE:notation}
We use the notations $\overline{\Omega}$ for the equilibrium value of
the order parameter and ${\Omega}$ for the fluctuating variable (which
features in the Landau free energy and of which $\overline{\Omega}$ is
the expectation value).
\bibitem{FNOTE:numerical_error} The numerical errors quoted are
estimated by using different numerical methods to compute the function
$\eta^{(0)}_{1}(x,x')$ and comparing the highest and lowest estimates
for the eigenvalues thus obtained. Other sources of error, not
considered in the quoted estimate, 
include the discretization of the integral equation and the numerical
error of the diagonalization algorithm itself.
\bibitem{FNOTE:zeta_nonzero} 
One could attempt to solve Eq.~(\ref{EQ:eqn_x_small}) by assuming that
$\zeta = 0$, but this leads to the condition 
$$
4 \langle \theta^{-1+nd/2} \rangle_{\pi} = l-1+nd/2,
\label{EQ:impossible}
$$
which can, at most, be satisfied for isolated values of $n$. In
particular, by using Eq.~(\ref{EQ:moment_value}) we see that for $n=0$
the left hand side is not an integer, and the equation is not
satisfied.
\bibitem{FNOTE:spurious_0rs} 
To examine this assertion, let us consider a Hilbert space slightly
different from the one defined in Sec.~\ref{SEC:H_change}, in which
the functions $f_{\hat{k}}$ are defined also for $\hat{k} =
\hat{0}$. This enlargement of the Hilbert space makes room for
spurious fluctuations of the order parameter at the origin to
appear. The scalar product in this enlarged space is defined by
$$
(f|g) 
\equiv \frac{1}{V^{n}} 
\sum_{\hat{k}} f^{\ast}(\hat{k}) \, g(\hat{k}).
\label{EQ:scalar_product_extend}
$$
The scalar product of the eigenvector ${\psi}^{s}$ associated with a radial
eigenfunction singular at the origin and the normalized function
${\psi}^{z}(\hat{k}) = V^{n/2} \delta_{\hat{k},\hat{0}}$ concentrated at the
zero replica sector is given by 
$$
({\psi}^{z} | {\psi}^{s}) = \frac{1}{V^{n/2}} {\psi}^{s}(\hat{0}) = 1
+ {\cal O}(n),
\label{EQ:conc_0rs}
$$
and thus ${\psi}^{s}$ lies entirely in the zero replica sector for $n
\to 0$.
\end{references}
\end{document}